\documentclass[preprint, nonacm]{acmart}
\usepackage{amsmath,amsfonts}
\usepackage{algorithmic}
\usepackage{booktabs} 
\usepackage{tabularx}
\usepackage{graphicx}
\graphicspath{{figure/}}
\usepackage{subcaption}
\usepackage{textcomp}
\usepackage{breqn}
\usepackage{hyperref}
\usepackage{siunitx}
\usepackage{xspace}
\usepackage{comment}
\usepackage{listings}
\usepackage{enumitem}
\usepackage{xcolor} 

\usepackage[font={small,sf,bf}]{caption}

\newif\iffinal

\iffinal
\newcommand{\raj}[1]{}
\newcommand{\zliu}[1]{}
\newcommand{\ian}[1]{}
\newcommand{\rachana}[1]{}

\else
\newcommand{\raj}[1]{{\textcolor{green}{ Raj: #1 }}}
\newcommand{\zliu}[1]{{\textcolor{blue}{ Zhengchun: #1 }}}
\newcommand{\ian}[1]{{\textcolor{red}{ Ian: #1 }}}
\newcommand{\rachana}[1]{{\textcolor{purple}{ Rachana: #1 }}}
\fi

\newcommand{\wasabi}{\texttt{Wasabi}\xspace}
\newcommand{\aws}{\texttt{AWS-S3}\xspace}
\newcommand{\boxcom}{\texttt{box.com}\xspace}
\newcommand{\gcs}{\texttt{Google-Cloud}\xspace}
\newcommand{\gdr}{\texttt{Google-Drive}\xspace}
\newcommand{\ceph}{\texttt{Ceph}\xspace}

\newcommand{\dsiwasabi}{\texttt{Wasabi-Connector}\xspace}
\newcommand{\dsiaws}{\texttt{AWS-Connector}\xspace}
\newcommand{\dsiboxcom}{\texttt{Box.com-Connector}\xspace}
\newcommand{\dsigcs}{\texttt{GoogleCloud-Connector}\xspace}
\newcommand{\dsigdr}{\texttt{GoogleDrive-Connector}\xspace}
\newcommand{\dsiceph}{\texttt{Ceph-Connector}\xspace}

\newcommand{\dsi}{DSI\xspace}
\newcommand{\gdsi}{\texttt{Connector}\xspace}

\renewcommand\footnotetextcopyrightpermission[1]{} 
\begin{document}

\title{Design and Evaluation of a Simple Data Interface for Efficient Data Transfer Across Diverse Storage}

\setcopyright{acmcopyright}
\copyrightyear{2020}
\acmYear{2020}
\acmDOI{10.1145/1122445.1122456}

\settopmatter{printacmref=false}

\author{Zhengchun Liu}
\affiliation{%
  \institution{Argonne National Laboratory}
  \streetaddress{9700 S. Cass Ave.}
  \city{Lemont}
  \state{IL}
  \postcode{60439}
}
\email{zhengchun.liu@anl.gov}
\orcid{1234-5678-9012}

\author{Rajkumar Kettimuthu}
\affiliation{\institution{Argonne National Laboratory}}
\email{kettimut@anl.gov}

\author{Joaquin Chung}
\affiliation{\institution{Argonne National Laboratory}}
\email{chungmiranda@anl.gov}

\author{Rachana Ananthakrishnan}
\affiliation{\institution{The University of Chicago}}
\email{rachana@globus.org}

\author{Michael Link}
\affiliation{\institution{The University of Chicago}}
\email{mlink@globus.org}

\author{Ian Foster}
\affiliation{\institution{Argonne National Laboratory}}
\affiliation{\institution{The University of Chicago}}
\email{foster@anl.gov}

\begin{abstract}
Modern science and engineering computing environments often feature storage systems of different types, from parallel file systems in high-performance computing centers to object stores operated by cloud providers. To enable easy, reliable, secure, and performant data exchange among these different systems, we propose \gdsi{}, a pluggable data access architecture for diverse, distributed storage. By abstracting low-level storage system details, this abstraction permits a managed data transfer service (Globus in our case) to interact with a large and easily extended set of storage systems. Equally important, it supports third-party transfers: that is, direct data transfers from source to destination that are initiated by a third-party client but do not engage that third party in the data path. The abstraction also enables management of transfers for performance optimization, error handling, and end-to-end integrity. We present the \gdsi{} design, describe implementations for different storage services, evaluate tradeoffs inherent in managed vs.\ direct transfers, motivate recommended deployment options, and propose a performance model-based method that allows for easy characterization of performance in different contexts without exhaustive benchmarking.
\end{abstract}


\maketitle

\section{Introduction}
Easy access to data produced by scientific research is essential if such research products are to be widely available for research, education, business, and other purposes~\cite{pasquetto2017reuse}. 
Far from being mere rehashes of old datasets, evidence shows that studies based on analyses of previously published data can achieve just as much impact as the original projects~\cite{data-share}.
Reducing barriers to the sharing of scientific data is a multi-faceted challenge~\cite{fair},
but one fundamental need is for mechanisms that can enable efficient, secure, and reliable access to data, regardless of location.

The work described here is concerned with addressing two important obstacles to
scientific data access, namely storage system diversity and efficient data movement.
Ideally, once a scientist locates data of interest, they should be able to retrieve required components easily, reliably, efficiently, and securely, without concern for the details of the source and destination storage systems.
In practice, 
considerations such as domain practices, cost, performance, data source, and analysis workflows result in scientists storing data on a wide range of storage systems, often with idiosyncratic interfaces: for example, commercial cloud object-based storage, such as Amazon Simple Storage Service (S3), Google Cloud Storage, and Microsoft Azure Blob Storage;
community object-based storage solutions, such as Ceph and OpenStack Swift;
parallel filesystems, such as Lustre, GPFS, and Intel DAOS;
and cloud-based file hosting services and synchronization services, such as Box, Google Drive, Microsoft OneDrive, and Dropbox.
Furthermore, the datasets that are created and analyzed in science are frequently large, reaching terabytes or even petabytes in size.
Thus the scientist must be able not only to access diverse data stores
but to optimize data movement among different combinations of such systems.
This need creates its own challenges in terms of leveraging these resources without overburdening application researchers~\cite{ssio}.

A first important step towards enabling seamless data access and movement among diverse storage systems was taken more than a decade ago, 
when Allcock et al.~\cite{globus-gridftp} introduced the Data Storage Interface (DSI) within the  Globus implementation of the GridFTP protocol as a unified storage interface for use by data movement tools. 
Globus GridFTP and its DSI were initially supported on POSIX-compliant file systems~\cite{globus-gridftp}, but the storage landscape increasingly includes cloud object stores, tape archives, and other proprietary storage systems. 
The evolution of DSI to accommodate new storage systems while maintaining backward compatibility, and also
to incorporate support for third-party transfers and modern authentication and authorization mechanisms, produced the \textbf{Connector} abstraction that we describe in this paper.
This abstraction, as instantiated in an interface and implementation,
enables a wide variety of storage systems to be
accessed in a consistent, performant, and secure manner, simply by installing the required \gdsi{} server software~\cite{globus}.
Equally important, it supports the management of transfers by cloud-hosted management services, such as the Globus service that we consider in this paper.

While a uniform interface to storage has many advantages,
any abstraction layer tends to introduce overheads that 
can impact performance.
Understanding the nature of these overheads is essential to determining where and when the abstraction may be used.  
To develop this understanding, we first present here the \gdsi{} abstraction and then evaluate overheads and performance when the Globus implementation of this abstraction is used to move data between diverse storage systems.
The primary contributions of this paper are the following:
\begin{itemize}[leftmargin=*]
\item We describe \gdsi{}, a data storage interface that permits uniform access to a wide range of storage systems, including both cloud storage and conventional file systems, while supporting third-party managed transfers.
\item We describe how this interface can be incorporated into a data transfer service.
\item We propose a performance model-based method for exploring performance issues in different contexts without exhaustive benchmarking.
\item We draw conclusions about implications for our design and the Globus implementation, and recommend best practices.
\end{itemize}

The rest of this paper is organized as follows. 
In \S\ref{sec:motivation} we describe the motivation for a uniform interface for data movement across diverse storage systems including the cloud based storage services.
In \S\ref{sec:dsi} we propose \gdsi{} based on the original DSI to address new challenges in cloud-based storage service.
The details of \gdsi{} and six sample implementations are introduced in \S\ref{sec:connector}.
In \S\ref{sec:overhead} we present a performance-model-based approach to study overhead of data movement using \gdsi{}, and in \S\ref{sec:throughput} we analyze the throughput of \gdsi{}-based data movement and in \S\ref{sec:integrity} we evaluate the influence of data integrity checking which shows unique characteristics to \gdsi{}.
The results presented in \S\ref{sec:throughput} are in line with the performance analysis in \S\ref{sec:overhead}.
In \S\ref{sec:best-practice} we discuss best practices for production deployment. 
In \S\ref{sec:related-work} we review related work , and in \S\ref{sec:conclusion} we summarize our conclusions and discuss future work. 

\section{Motivation}\label{sec:motivation}
The science and engineering community uses a large and growing number of storage systems. Each such system has been created in response to specific needs for storing and accessing science and engineering data, 
needs that cover a broad spectrum of cost, scale, performance, and other requirements.
Different systems focus on distinct requirements, provide distinct services to their clients, and often implement different interfaces and protocols for data access. 

For example, POSIX I/O provides open(), close(), read(), write(), and lseek() operations, with strict consistency and coherence requirements. 
For instance, write operations have to be visible to other clients immediately after the system call returns. 
POSIX serves as the uniform interface for many storage systems, such as Lustre and GPFS, that are widely used in science institutions.
Although high-level parallel I/O middleware libraries, such as MPI-IO~\cite{thakur1999implementing}, HDF5~\cite{folk1999hdf5}, ADIOS~\cite{liu2014hello}, and PnetCDF~\cite{1592942}, provide relaxed semantics for file system access, most scientific HPC applications use POSIX as their default interface for interacting with local storage~\cite{ics20-alcf-logs}. 

Object stores, widely used in cloud computing, manage data as objects instead of files. 
They provide a single flat global name space, support just a few simple operations, such as PUT and GET, and support a weak form of consistency: eventual consistency.
Although APIs provided by cloud storage providers enable high-speed access to data from \emph{within} the cloud, they do a bad job of moving data \emph{between} cloud and local science institutions or among different cloud service providers. 
The user must log into the system in order to perform download or upload operations between cloud storage and another institution.  
This data movement pattern is unreliable, since any system interruption results in failure, and is difficult to incorporate into research automation tools such as workflow systems and HPC schedulers, which typically use POSIX interface for data staging.


\subsection{Third-Party Transfer}\label{sec:3rd-party}
An important data transfer pattern in many science and engineering settings involves a ``third party'' (a user or agent working on behalf of a user) initiating and managing data movement between two remote computers (or instruments).
Support for such \emph{third-party data transfers} allows users to initiate, manage, and monitor data movement from anywhere, without direct access to the systems involved in the data movement.
It also facilitates the integration of data movement operations into a wide variety of data automation tools, from shell scripts to
scientific workflow engines,
such as Galaxy~\cite{goecks2010galaxy}, Kepler~\cite{Kepler2004},
Parsl~\cite{babuji19parsl},
Pegasus~\cite{deelman2015pegasus},
and Swift/T~\cite{wozniak2013swift}.
The ability to request data transfers enables the workflow systems to
execute transparently on remote resources.

A third-party transfer necessarily engages two distinct communication channels, 
for control and data.  
The control channel is used for sending protocol messages between system components, for example, from the third-party management service to data movers, in order to authenticate and authorize users and to initiate streaming. The data channel provides the link between the source and destination of the data  being transferred. 
Cloud-based storage services, in contrast, provide only two-party (i.e., server-client) data movement, which makes it difficult to integrate such services into scientific workflows in which some components may run on supercomputers that do not have WAN connectivity and typically communicate with dedicated data transfer nodes via parallel file systems. 

\subsection{Transfer Management}

While the servers connected to storage are the workhorses for data movement, the clients that drive data transfers play a critical role in determining transfer performance and reliability. 
The client 
needs to provide all parameters for any transfer, including the security credential to be used, network usage levels (e.g., number of concurrent connections, parallelism), and integrity and privacy levels. 
The configuration of these parameters by the client, as part of the request that it issues to servers to transfer files, has significant impact on transfer performance. Moreover, while the GridFTP protocol is built for supporting reliable transfers, the onus falls on the client to track the information sent back from the servers on how much data has been moved and request restarts when these are needed to ensure a complete transfer. Similarly, any failures, including those that result from integrity checks performed upon completion of a file transfer, are returned to the client; it is up to the client application to request that data be retransferred. For large transfers that include recursive transfers of directories, the client needs to expand directories and track progress at a per-file level in order to ensure that all files and folders are moved. 

Client applications that provide such transfer management capabilities and the rich features required to drive high performance and reliability are not trivial to write and maintain. The Globus transfer service provides such a client as a hosted service for managed transfers. It drives maximum efficiency by combining information on the files that need to be moved and the capacity at source and destination to determine performance parameters. It also provides reliability by tracking transfer progress and retrying on faults, and it  negotiates the security needed to navigate transfers between sites. By thus providing a fire-and-forget solution for users, it delivers significant usability benefits.

The \gdsi{} storage abstraction layer thus has two roles: (1) to enable efficient access to data stored on a variety of different storage systems and (2) to support the operation of managed transfer applications such as Globus so that they can achieve secure, performant, and reliable transfers across diverse combinations of such systems.
The latter role requires 
specialized capabilities, such as enabling a third party to establish their identity and authority to make a request to the storage system; request that a transfer be initiated; enable encryption; monitor transfer progress; and detect errors and termination; request checksums.

\section{A uniform Storage Interface: from DSI to Connector} 
\label{sec:dsi}
\begin{figure}[htb]
\centering
\includegraphics[width=0.7\linewidth]{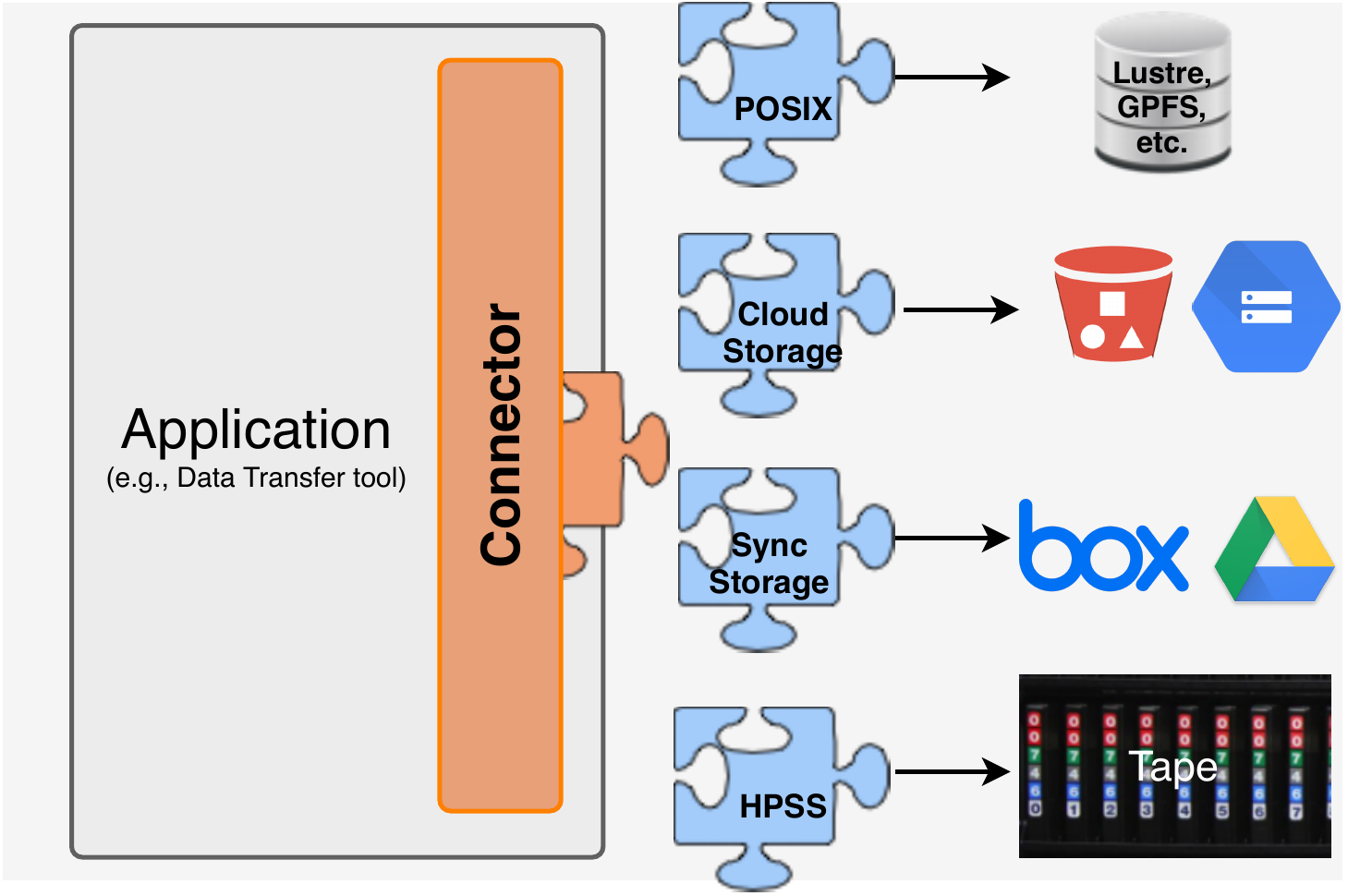}
\caption{The \gdsi{} abstraction}
\label{fig:dsi}
\end{figure}

Allcock et al.~\cite{globus-gridftp} described the initial DSI~(illustrated in \autoref{fig:dsi}) in 2005 targeting a uniform interface for various \textit{local} storage systems and data management systems such as HPSS~\cite{hpss}, Xrootd~\cite{xrootd}, and iRODS~\cite{irods}. 
This original DSI provides a uniform interface to the data storage systems for file transfers between local storage systems; it was intended primarily for use by the the open source Globus implementation of GridFTP.  
The DSI layer accepts requests such as stat, send, and recv and performs these functions using the appropriate APIs of the storage system with which it interfaces. 
DSI consists of several function signatures and a set of semantics. 

Continued advances and diversification in network bandwidth and data storing/management techniques, such as object stores, cloud-based storage services, and conventional parallel file systems, led to the original DSI design being no longer able to handle efficiently the many different data store/retrieval APIs and diverse authentication methods.  
Subsequent extensions to the original DSI to support modern authentication methods, handle certain limitations of cloud storage APIs (such as call quotas), incorporation of automatic retries and fault-tolerant capabilities, and additional management capabilities, produced what we refer to here as the \gdsi{} abstraction, as shown in \autoref{fig:gdsi} and detailed in \S\ref{sec:connector}.
An implementation of this abstraction is created by instantiating the various \gdsi{} functions. 
The first \gdsi{} developed by the Globus team was for POSIX-compliant file systems. Since then, researchers around the world have implemented  others. Examples include HPSS~\cite{hpssdsi}, iRODS~\cite{b2stage}, StoRM~\cite{stormdsi}, SDQuery~\cite{su2013sdquery}, Xrootd~\cite{dorigo2005xrootd}, Swift~\cite{swift-dsi}, and MAPFS~\cite{PEREZ2006620}---some in collaboration with the Globus team and some independently. To set context,  we introduce key \gdsi{} interface functions:
\begin{itemize}[leftmargin=*]
    \item \underline{Start} is called to establish a new session to access the storage. This hook gives a \gdsi{} an opportunity to
 set internal state that will be threaded through to all other function calls associated with this session.  It also provides an opportunity to reject the access request.
 \item \underline{Destroy} is called to terminate a session. The \gdsi{} should clean up all memory associated with the session. 
  \item \underline{Stat} is called to get information (e.g., size, last modified time) about a given file or resource and to verify that a file exists and has the proper permissions.
  \item \underline{Command} handles simple (succeed/fail or single-line response) storage system operations such as directory or object creation and  permission changes.  
  \item \underline{Recv} is used to receive data from the application and write to the underlying storage system.
  \item \underline{Send} is used to read data from the underlying storage system and send the data to the application.
  \item \underline{SetCredential} allows the application to provide the credential required for the \gdsi{} to authenticate with the underlying storage system. 
\end{itemize}

\begin{figure}[htb]
\includegraphics[width=.9\linewidth,trim=0cm 0mm 0cm 13mm,clip]{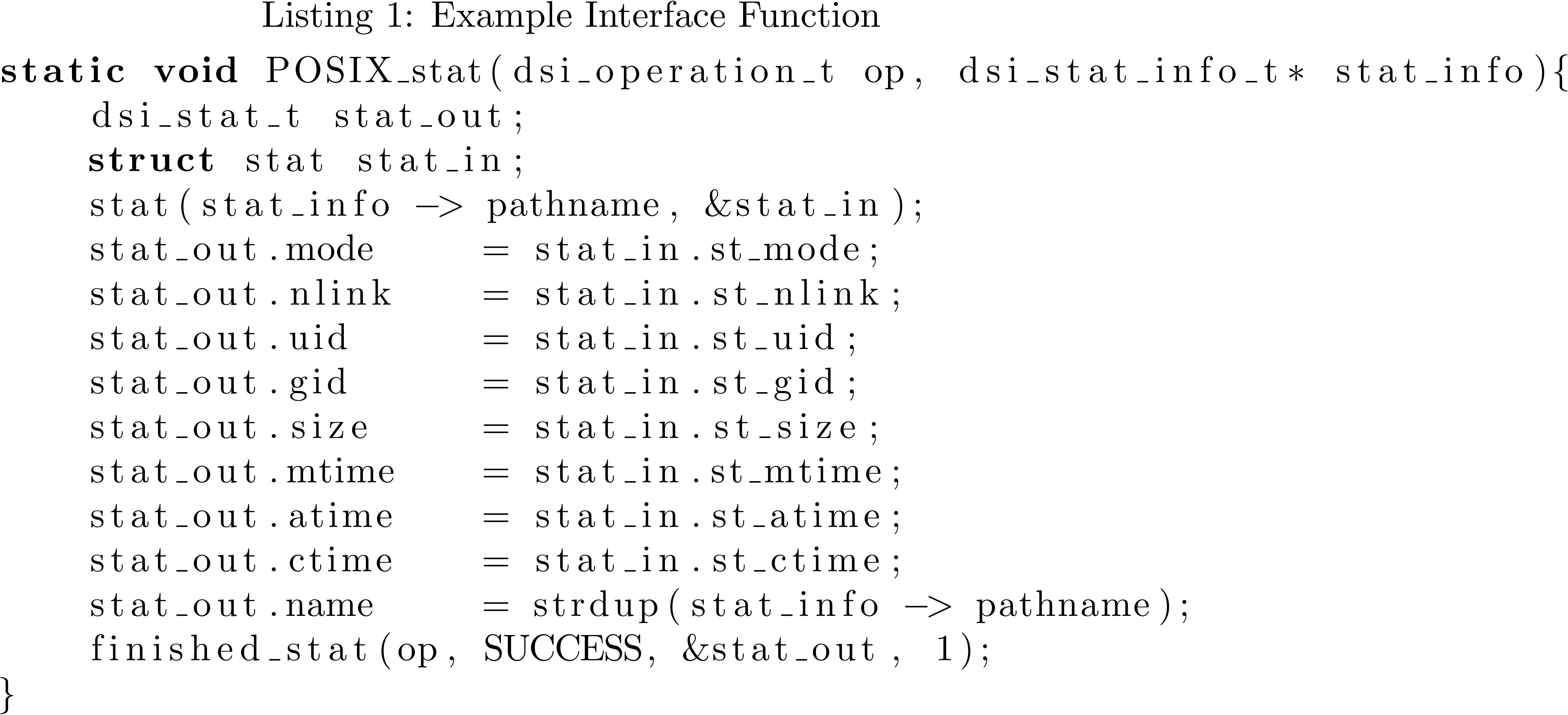} 
\caption{The \texttt{stat()} interface function as implemented for the POSIX Connector.}
\label{fig:stat-code-sample}
\end{figure}

In addition to these interface functions, the \gdsi{} includes various helper functions.
For example, 
\autoref{fig:stat-code-sample} shows the implementation of the \texttt{stat()} function for the POSIX \gdsi{}.  
The \textit{finished\_stat} call at the end of the function above is a helper function (implemented by the application) that allow the \gdsi{} to interact with the application. Here is a list of other helper functions:
\begin{itemize}[leftmargin=*]
    \item \underline{read/write} transfers data between \gdsi{} and application.
    \item \underline{get\_concurrency} tells the \gdsi{} how many outstanding reads or writes it should have. A data transfer application would specify this value based on the number of parallel TCP streams used for the data transfer.
    \item \underline{get\_blocksize} indicates the buffer size that the \gdsi{} should exchange with the application via read/write.
    \item \underline{get\_read\_range} tells the \gdsi{} which data it should be sending. This handles restart (including ``holey'' transfers) and partial data transfers.
    \item \underline{bytes\_written} should be called whenever the \gdsi{} successfully completes a write to the storage system. This allows the data transfer application to generate performance and restart markers.
\end{itemize}

Applications can load and switch \gdsi{} at runtime.
When the application requires action from the storage system (e.g., store/retrieve data, metadata, directory creation), it passes a request to the loaded \gdsi{} module. 
The \gdsi{} then services that request and notifies the server when it is finished. 
An API is provided to the \gdsi{} author to assist in implementation. 
The \gdsi{} author is not expected to know the details of the application. 
Instead, this API provides functions for reading and writing data to and from the network. 

\section{Globus Architecture and Connector Implementation}\label{sec:connector}
\begin{figure}[htb]
\centering
\includegraphics[width=0.8\linewidth]{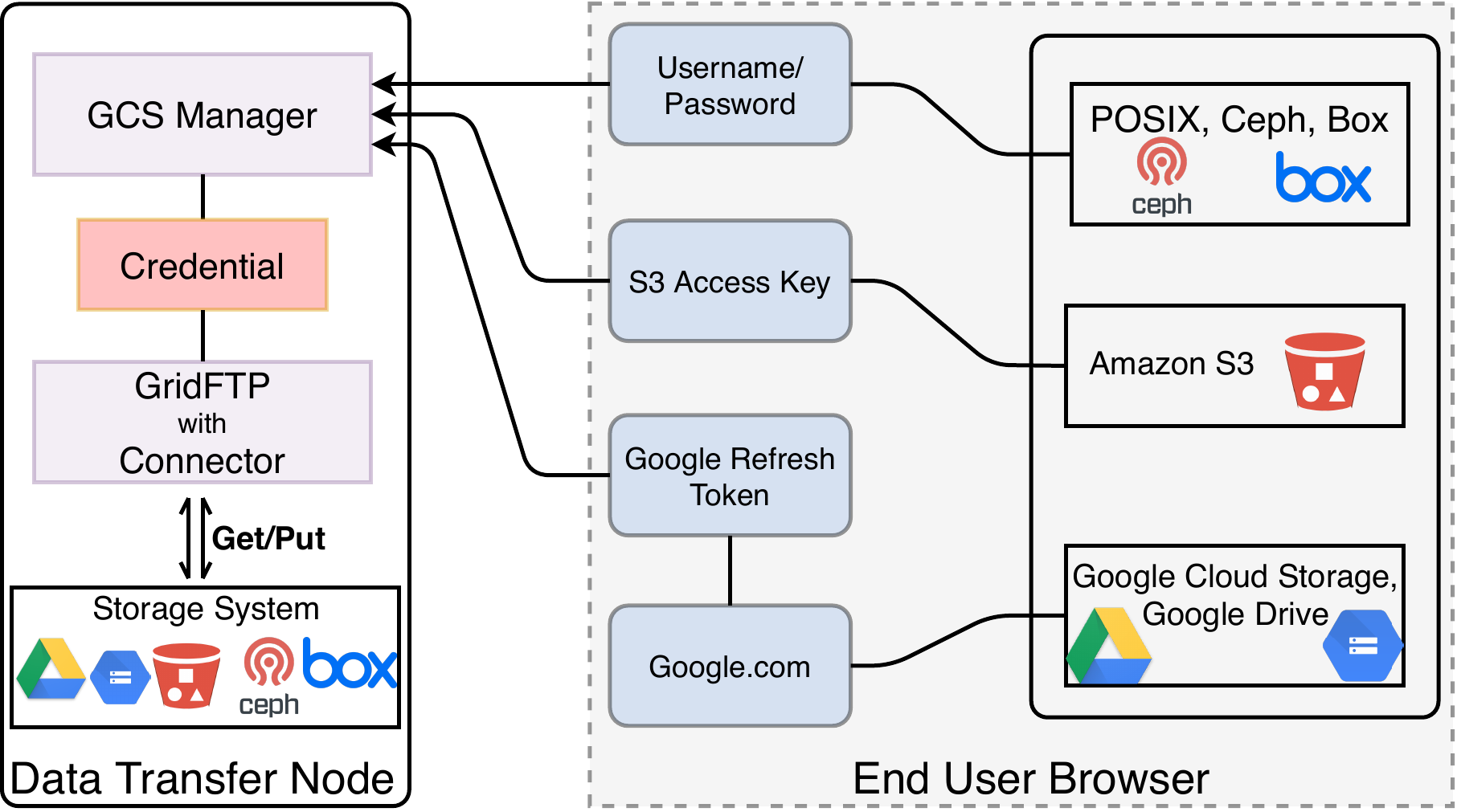}
\caption{Data- and Authentication- flow of \gdsi{}.}
\label{fig:gdsi}
\end{figure}

Rapid growth in both the use of the Globus service and the variety of available cloud and other storage resources has created the need for additional \gdsi{}s. 
We describe here six popular \gdsi{}s that are integrated into the GCS~\cite{globus}.
GCS comprises components for data access, security, and installation and configuration. 
For bulk data access, GridFTP---a high-performance, secure, reliable data transfer protocol optimized for high-bandwidth wide-area networks---is used.
GCS also includes a HTTPS server for direct two-party data access. 
The storage \gdsi{} provide the abstract layer for secure access to a storage type and include key capabilities such  as security protocols needed by the storage system, management of security credentials or tokens, limit and throttling policy management, and the key IO access to the storage system. 
\autoref{fig:gdsi} shows the flow of authentication to make a \gdsi{} work with Globus. 

Every storage blob 
requires that a credential be registered with the endpoint's GCS Manager service using its REST API. 
The credentials are never sent via the hosted Globus transfer service; instead, they are sent directly from the user's client (browser or command line client) to the GCS server. 
When the storage blob is accessed, the credential is read from the GCS Manager by the GridFTP server and passed to the \gdsi{}.
More specifically for POSIX, Box, and Ceph connectors, the credential is simply the local username to which the user's login identity is mapped.
For AWS S3, the credential is a user-submitted S3 Access Key ID and Secret Key. 
For Google Drive and Google Cloud Storage, the credential is a token that is sent to the GCS Manager directly by Google, after the user completes the Google OAuth2 login.
Non-persistent GridFTP clients such as
\texttt{globus-url-copy}~\cite{globus-gridftp} pass through credentials; users must provide them anew upon each failure. 
In the rest of this section, we briefly introduce the cloud stores with which we evaluate \gdsi{} later using a performance modeling based method, one of our key contributions of this paper.

Amazon Simple Storage Service (\aws{}) is a service offered by Amazon Web Services (AWS) that provides object storage through a web service interface.

\wasabi{}~\cite{wasabi}, an enterprise-class, tier-free cloud storage service, 
provides an S3-compliant interface to use with storage applications, gateways, and other platforms.

\gdr{} is a file storage and synchronization service developed by Google.
G Suite~\cite{gsuit}, a suite of cloud computing, productivity and collaboration tools, software, and products developed by Google Cloud, is widely used by education institutions, with which users have significant storage allocations with Google Drive. It also is being used as second-tier storage for research data. \gdsi{} helps in handling certain limitations of the Google Drive API (such as call quotas) through automatic retries and fault-tolerant capabilities in Globus transfer service. 

\ceph{}~\cite{weil2006ceph}, an open-source software storage platform, 
delivers object, block, and file storage in one unified system. It is based on Reliable Autonomic Distributed Object Store, which distributes objects across a cluster of storage nodes and replicates objects for fault tolerance. Ceph decouples data and metadata. Object Storage Devices store data, and Metadata Servers stores metadata, with metadata distributed dynamically among multiple Metadata Servers.  

\boxcom{} provides a  service similar to \gdr{}. 
Its growing use in universities and national laboratories for research data
motivated the development of a Box \gdsi{}, which enables bridging to other storage and, as with Google Drive, handles limitations of the native API. 

\gcs{} storage, like \aws{}, is a RESTful file storage web service for storing and accessing data on Google Cloud Platform infrastructure. 
Its service specifications make it more suitable than \gdr{} for enterprise use.
Its growing use for research data motivated the implementation of a \gcs{} \gdsi{}, which is being used to move data between \gcs{} and research institute storage and between \aws{} and \gcs{}.

\section{Performance Modeling Based Overhead Evaluation}\label{sec:overhead}
Liu et al.\ observe that the per-file overhead is the performance killer when transferring many small files between science facilities~\cite{ccgrid-19}, and that science workloads often have that unfortunate characteristic~\cite{hpdc18-zliu}.
Although the influence of per-file overhead can be alleviated by transferring many files concurrently, the DTN resource requirement will be higher to support large concurrency (the number of files transferred concurrently). 
We describe here a performance model that captures per-file overheads, and we present experiments that allow us to measure indirectly the per-file overhead when \dsi{} cloud \gdsi{} are involved in a transfer.
All experiment source code, environment setup instructions and experiment result analysis code are available at \url{https://github.com/ramsesproject/dsi}.

\subsection{Experiment Design}\label{sec:doe}
\begin{figure}[htb]
\centering
\includegraphics[width=.9\linewidth]{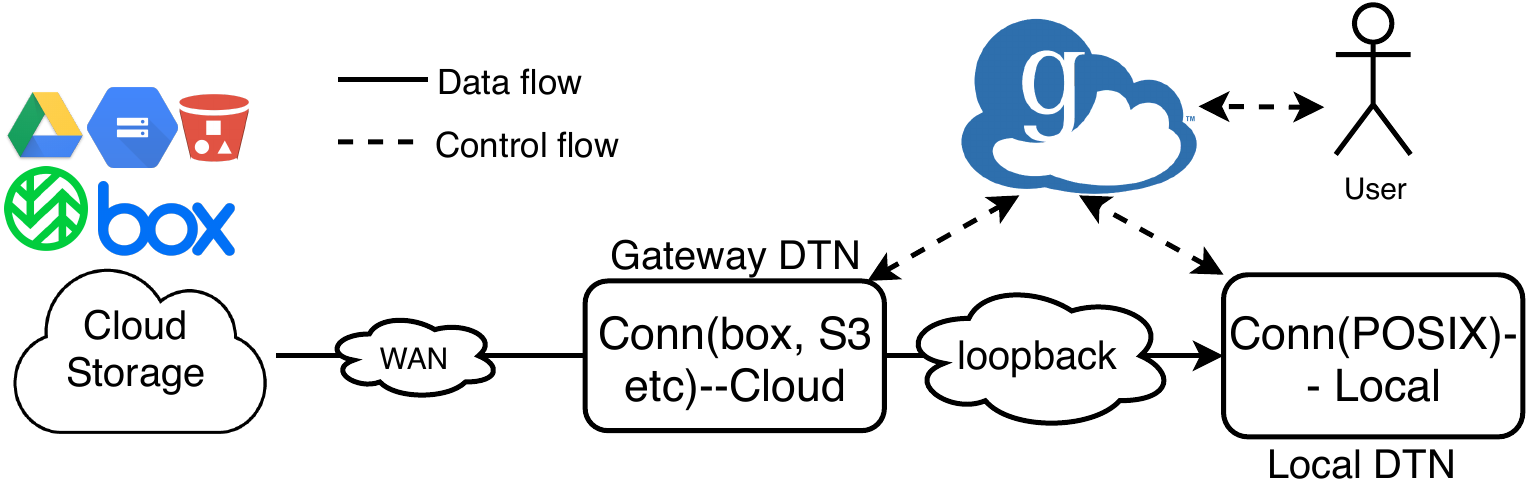}
\caption{A typical scenario in which the \gdsi{} is deployed locally in the science institution.}
\label{fig:dsi-gateway-all}
\end{figure}

\autoref{fig:dsi-gateway-all} shows a configuration in which the \gdsi{} is deployed on a DTN managed by a research institution. We refer to this configuration as \texttt{Conn-local}. 

\begin{figure}[htb]
\centering
\includegraphics[width=.7\linewidth]{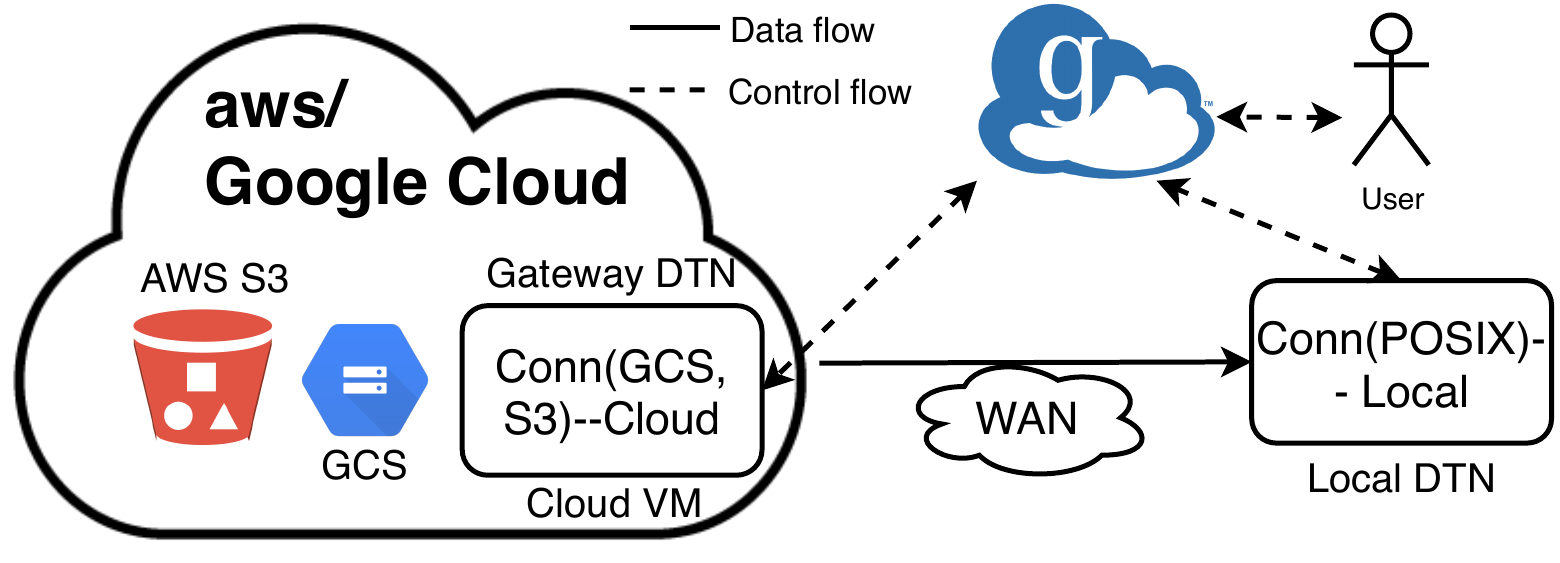}
\caption{A scenario in which the \gdsi{} is deployed in the same cloud as storage, as seen in \aws{} and \gcs{}.}
\label{fig:dsi-incloud}
\end{figure}

GridFTP~\cite{globus-gridftp} has been optimized for moving data efficiently over wide area network~\cite{pb-a-day}. 
Thus, for \aws{} and \gcs{}, we evaluate another deployment scenario in which the \gdsi{} runs on a VM operated by the same cloud provider, ideally in the same region as the storage. 
\autoref{fig:dsi-incloud} shows the case where the GCS and corresponding storage \gdsi{} are deployed in the same region as the cloud storage. 
Here, cloud storage (\aws{} or \gcs{}) APIs are used only for local data access and GridFTP is used to move data over the wide area network. 
We refer to this configuration as \texttt{Conn-cloud}.

\subsection{Analysis of Experiment Results}
\subsubsection{Regression analysis}\label{sec:reg-mdl}
These statistical processes enable estimation of the relationships between a dependent variable (e.g., data movement performance) and one or more independent variables (e.g., file sizes). 
Consider the model function 
\begin{equation}
y=\alpha +\beta x,
\label{eq:linear-mdl}
\end{equation}
which describes a line with slope $\beta$ and y-intercept $\alpha$. 
This relationship may not hold exactly for the largely unobserved population of values of the independent and dependent variables; we call the unobserved deviations from this equation the errors. 
Suppose we observe $n$ data pairs and call them {($x_i$, $y_i$), $i$ = $1, \ldots n$}. 
We can describe the underlying relationship between $y_i$ and $x_i$ involving the error term $\epsilon_{i}$ by 
\begin{equation}
Y_{i} = \alpha + \beta x_{i}\textbf{} + \epsilon_{i}.
\label{eq:reg-mdl}
\end{equation}

We can then estimate $\alpha$ and $\beta$ by solving the following minimization problem:
\begin{equation}
\min_{\alpha, \beta }Q(\alpha, \beta ),\quad {\text{for }}Q(\alpha, \beta )=\sum _{i=1}^{n}(y_{i}-\alpha -\beta x_{i})^{2}.
\label{eq:lr-solution}
\end{equation}

\subsubsection{Performance model}\label{sec2:perf-mdl}
We consider the transfer of multiple files sequentially between two endpoints. 
We assume that each file introduces a fixed overhead of $t_0$~\cite{ccgrid-19}, and the end-to-end theoretical throughput (the minimum of source read, network, and destination write, as studied by Liu et al.~\cite{hpdc17-zliu}) is \emph{R}. Then 
the time \emph{T} to transfer \emph{N} files totaling \emph{B} bytes using a concurrency of one (i.e., transfer \emph{N} files one-by-one) is
\begin{equation}
T=N \times t_{ 0 }+\frac { B }{ R } + S_0,
\label{eq:trs-mdl}
\end{equation}
where $S_0$ is the transfer startup cost in seconds (to be measured in \S\ref{sec:startup-cost}).
$S_0$ will be close to zero for two-party transfer (e.g., using native API) but higher for third-party (e.g., cloud-hosted Globus) transfer because there will be coordination cost between transfer client and servers at source and sink.  

We then use \autoref{eq:trs-mdl} to measure indirectly the per-file overhead using \autoref{eq:lr-solution} by letting $\alpha=\frac { B }{ R } + S_0$ and $\beta=t_0$.
Since $S_0$ typically is 
a constant value, $\alpha$ will reflect the \texttt{network use efficiency}, namely, how fast the network can transfer one single large file. 

\subsubsection{Pearson correlation}
This coefficient~\cite{benesty2009pearson}, $\rho\left(x, y\right)$, is a measure of the linear correlation between variables $x$ and $y$. 
It has a value between +1 and $-$1, where 1 is total positive linear correlation, 0 is no linear correlation, and $-$1 is total negative linear correlation.
It 
is calculated by:
\begin{equation}
\rho\left(x, y\right) = \frac{\text{cov}\left(x, y\right)}{\sigma_x\sigma_y}, 
\label{eq:pearson}
\end{equation}
where $\text{cov}\left(x, y\right)=E\left[(X-\mu_x)(Y-\mu_y)\right]$ is the co-variance of variables $x$ and $y$, and $\sigma_x$ and $\sigma_y$ denote the standard deviation of $x$ and $y$, respectively. 

We performed experiments to verify this model and to estimate the overhead (i.e., $t_0$) of a file indirectly.
We then used Pearson correlation to quantify the linear relation between data transfer time, \emph{t}, and number of files, \emph{f}, in the datasets that total to the same size across all experiments.
\autoref{tbl:cc} presents the correlation coefficient between data transfer time and number of files for transfers to and from the six storage systems using \gdsi{} (deployed locally and at cloud if applicable) as well as native API. 
We see that the coefficients are close to 1 for all cases, indicating a strong linear relation between transfer time and number of files. 
Thus we can use \autoref{eq:trs-mdl} as a performance model for data transfers to/from all six storage systems and use regression analysis to resolve the model parameters. 
 
In all experiments, we kept the total dataset size fixed but varied the number of files. 
In choosing dataset sizes, we aimed to keep the experiment time short enough to reduce the influence of fluctuating external load such as storage and network, but long enough to mitigate transfer service startup cost (i.e., per-transfer overhead).
To this end, we used a 5~GB dataset for \wasabi{}, \aws{}, and \gcs{} and a 1~GB dataset for \gdr{} and \boxcom{}, considering their difference in peak end-to-end performance. 
We split each dataset into $N\in$ \{50, 100, 200, 400, 600, 800, 1000\} equal-sized files.

We moved the dataset using both the appropriate \gdsi{} and the service providers' native APIs and then used the regression analysis methods described above to build a performance model for both the \gdsi{} and the native API.
To mitigate the influence of external load on the cloud and network, we repeated each experiment between 3 and 10 times, depending on the observed fluctuation.

\begin{table}[htb]
\centering
\caption{Correlation coefficient, $\rho_{\left(t,f\right)}$, between transfer time \emph{t}, and number of files, \emph{f}, to and from different storage systems using the native API and \gdsi{} deployed locally (at a science institution) and at the cloud (where applicable).}
\begin{tabular}{l|c|c|c}
\noalign{\hrule height 2pt}
Transfer Direction & \gdsi{}-Local & \gdsi{}-Cloud & Native-API \\\hline
To \aws{} & 0.999 & 0.973 & 0.995\\\hline
From \aws{} & 0.989 & 0.993 & 0.989 \\\hline
To \wasabi{} & 0.999 & N/A & 0.998\\\hline
From \wasabi{} & 0.997 & N/A & 0.998 \\\hline
To \gcs{} & 0.997 & 0.999 & 0.993\\\hline
From \gcs{} & 0.999 & 0.996 & 0.992 \\\hline
To \gdr{} & 0.994 & N/A & 0.992\\\hline
From \gdr{} & 0.989 & N/A & 0.995 \\\hline
To \ceph{} & 0.996 & 0.999 & 0.999\\\hline
From \ceph{} & 0.986 & 0.994 & 0.976 \\\hline
To \boxcom{} & 0.998 & N/A & 0.999\\\hline
From \boxcom{} & 0.996 & N/A & 0.998 \\
\noalign{\hrule height 2pt}
\end{tabular}
\label{tbl:cc}
\end{table}

\subsection{Results and Discussion}
We performed the following experiments without Globus integrity checking~\cite{globus-online,globus}. Globus integrity check detects any data corruption that occurred during transmission over the network and/or while writing data to the destination storage by reading the data at the destination (after it was written to the storage), computing the checksum, and verifying it with the source checksum.  
In practice, integrity checking is of vital importance for storage-to-storage transfers~\cite{stone2000crc,hpdc18-zliu,charyyev2019towards}.
See \S\ref{sec:integrity} for more discussion of the importance and performance impacts of integrity checks.

\subsubsection{Amazon S3}
\begin{figure}[htb]
\centering
\begin{subfigure}[h]{.49\columnwidth}
\includegraphics[width=\columnwidth,trim=2.5mm 2mm 2mm 2mm,clip]{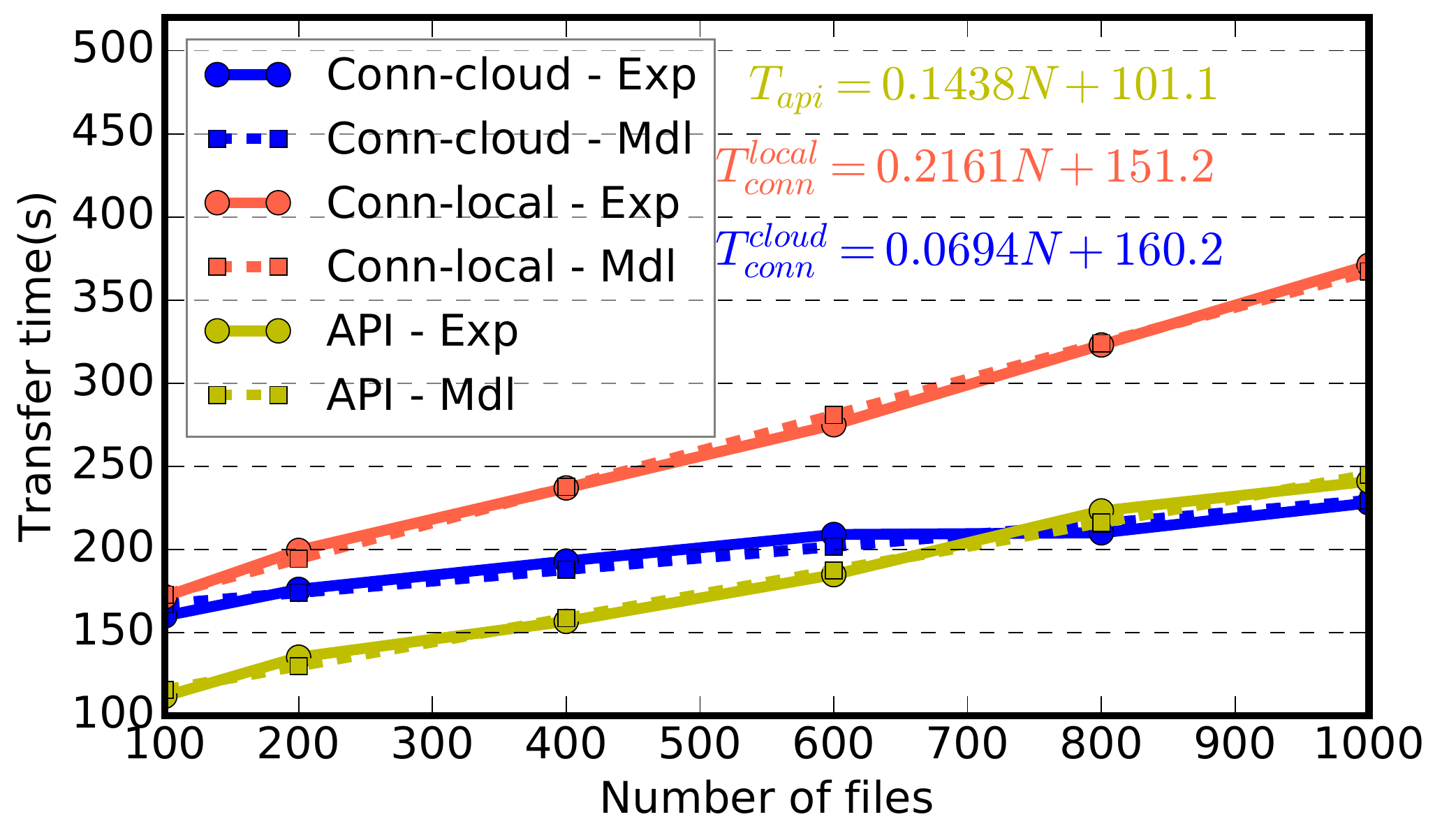}
\caption{Upload to \aws{}}
\label{fig2:aws-reg-ana-up}
\end{subfigure}
\begin{subfigure}[h]{.49\columnwidth}
\includegraphics[width=\columnwidth,trim=2.5mm 2mm 2mm 2mm,clip]{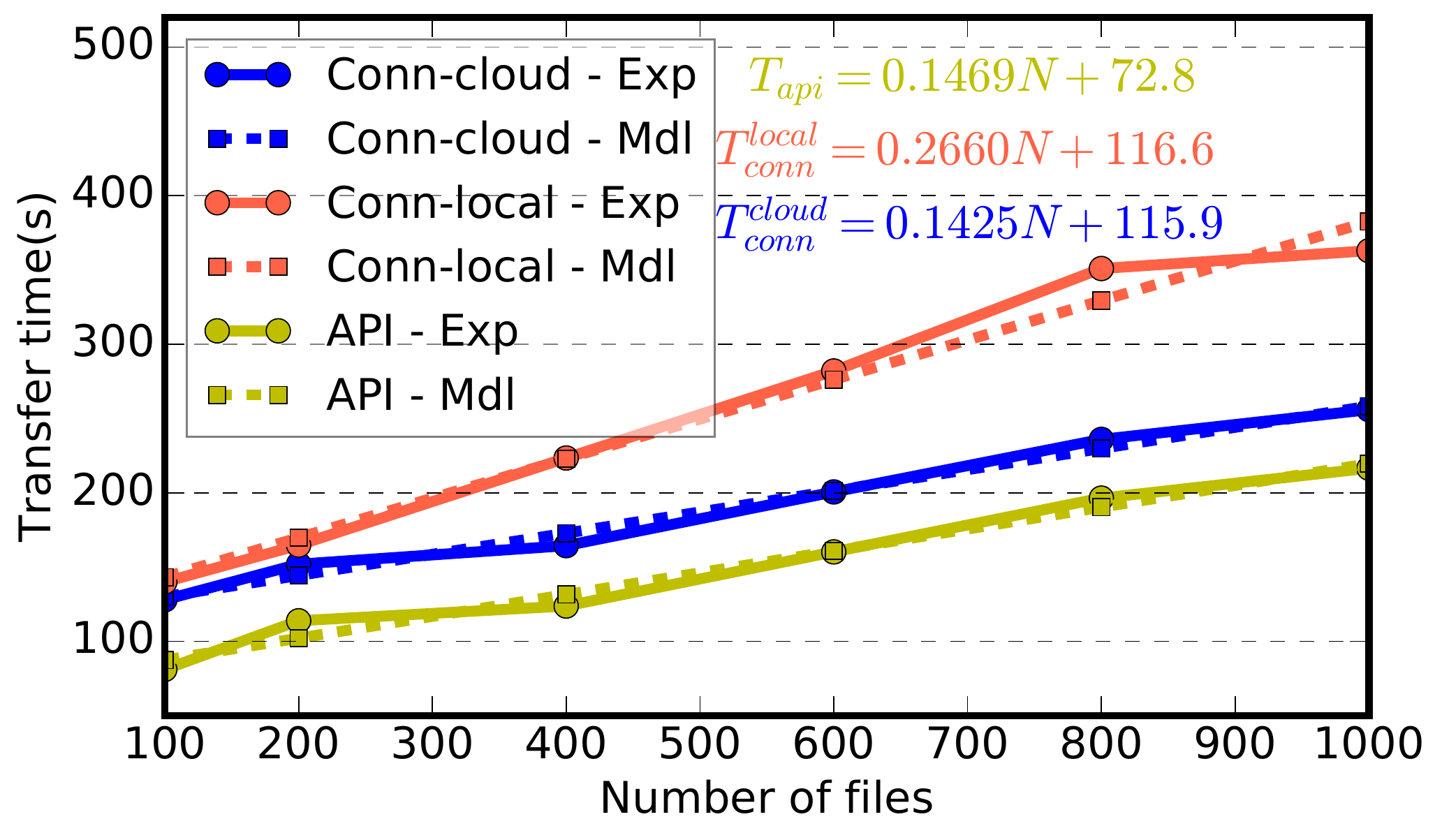}
\caption{Download from \aws{}}
\label{fig2:aws-reg-ana-down}
\end{subfigure}
\caption{Transfer time for 5~GB between local file system and \aws{} vs.\ number of files. \texttt{Conn-cloud}: \dsiaws{} is deployed in AWS near the S3 bucket~(\autoref{fig:dsi-incloud}); \texttt{Conn-local}: deployed in a science institution~(\autoref{fig:dsi-gateway-all}).}
\label{fig:aws-reg-ana}
\end{figure}

We used \texttt{boto3}~\cite{boto3}, a Python interface to AWS, as the native tool to download from and upload to \aws{} buckets, and we compared its performance with that of \dsiaws{}.
\autoref{fig:aws-reg-ana} shows the experimental results and performance model predictions. 
When uploading from local to \aws{}, \texttt{Conn-local} performs worse than native API; \texttt{Conn-cloud} has less per-file overhead but lower throughput than the native API has.
Thus, \gdsi{} can outperform native API when transferring many small files, where per-file overhead become significant.
For downloads from \aws{} to local file system, the per-file overhead trend is similar to that for uploads, but \dsiaws{} efficiency is worse when compare with native APIs, leading to worse performance when downloading large files. 

\subsubsection{Wasabi}
\wasabi{} provides an S3-compliant interface to use with storage applications, gateways, and other platforms.
We used APIs from \texttt{boto3} (the same API used for \aws{}) as the native tool to download from and upload to \wasabi{} buckets, and we compared its performance with the \wasabi{} connector.

\begin{figure}[htb]
\centering
\begin{subfigure}[h]{.49\columnwidth}
\includegraphics[width=\columnwidth,trim=2.5mm 2mm 2mm 2mm,clip]{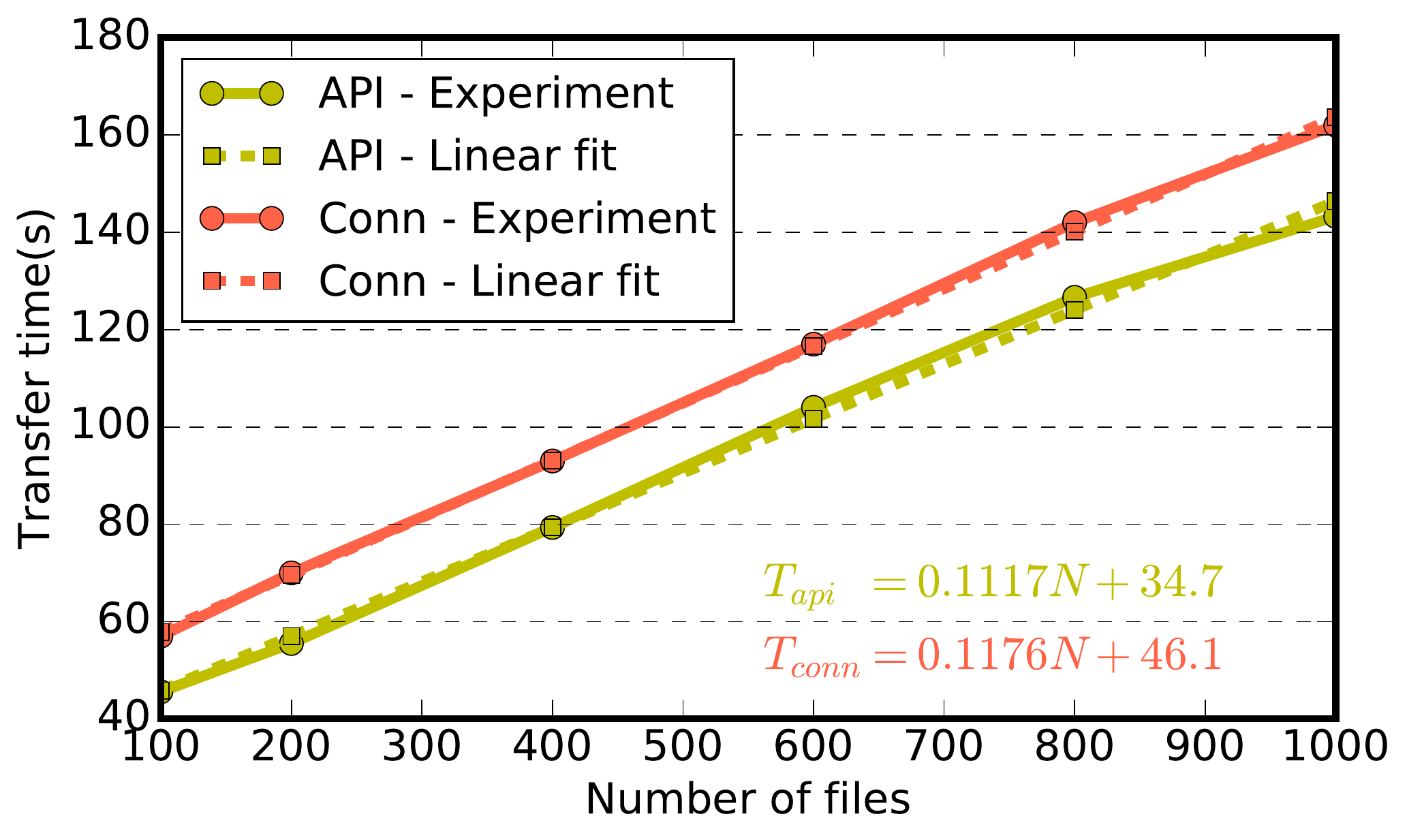}
\caption{Upload to \wasabi{}}
\label{fig2:wasabi-reg-ana-up}
\end{subfigure}
\begin{subfigure}[h]{.49\columnwidth}
\includegraphics[width=\columnwidth,trim=2.5mm 2mm 2mm 2mm,clip]{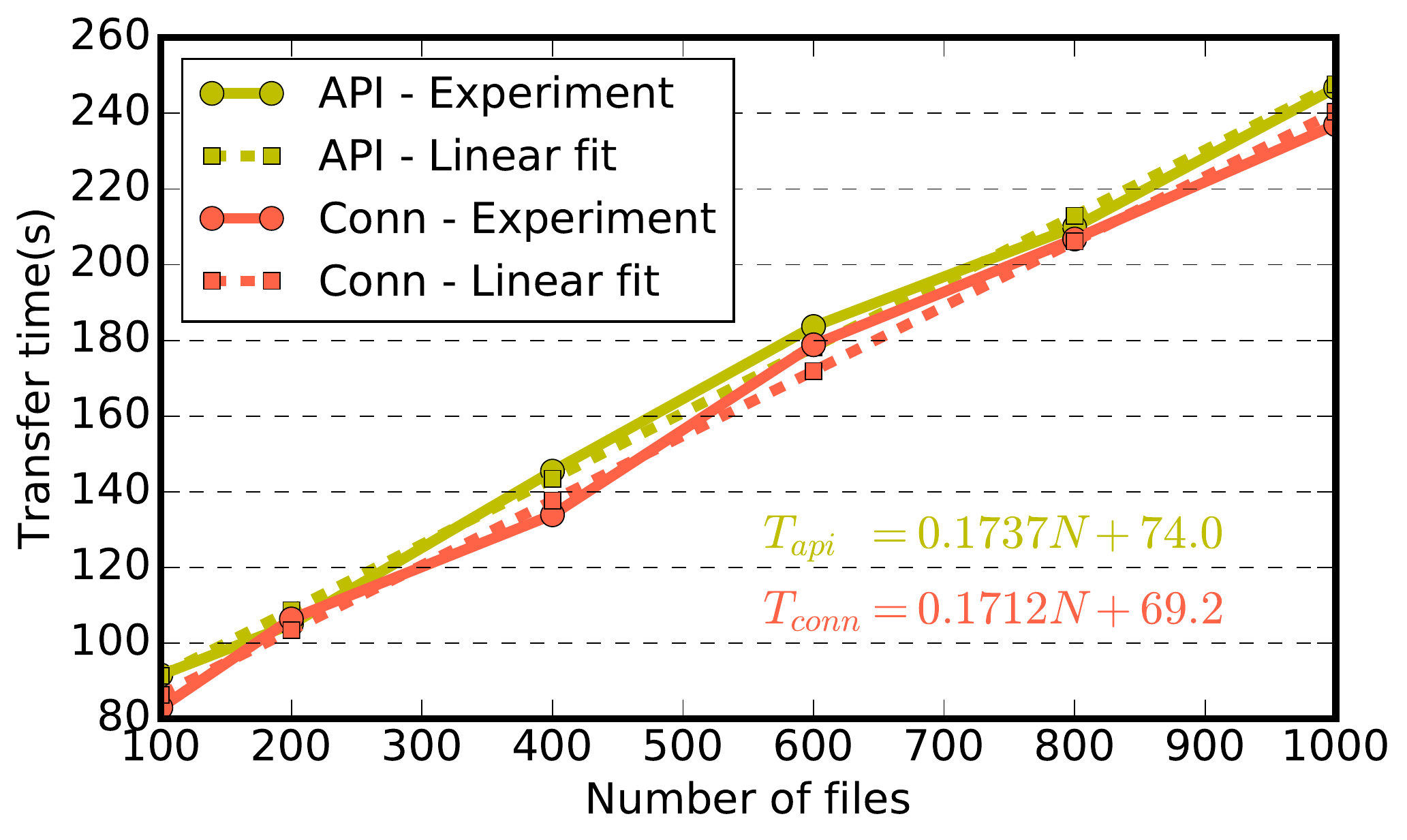}
\caption{Download from \wasabi{}}
\label{fig2:wasabi-reg-ana-down}
\end{subfigure}
\caption{Transfer time vs.\ number of files for 5~GB between local filesystem and \wasabi{}}
\label{fig:wasabi-reg-ana}
\end{figure}

\autoref{fig:wasabi-reg-ana} presents the regression analysis results for both upload to and download from \wasabi{} using \texttt{boto3} and the Globus \wasabi{} connector.
We see from \autoref{fig:wasabi-reg-ana} that the native tool and 
the \gdsi{} have similar per-file overheads, for both download from- and upload to \wasabi{}.
In terms of average throughput achieved, the \gdsi{} is slightly slower for uploads and slightly faster for downloads. 
Overall, we conclude that the \gdsi{} will perform worse than the native tool when uploading many large files, but it will be comparable to the native tool when transferring many small files, a common use case in practice.

We note that this experiment was designed to measure per-file overhead. This overhead can be mitigated by using either high concurrency or prefetch, as studied by Liu et al.~\cite{ccgrid-19}. We study throughput performance in 
\S\ref{sec:throughput}.

\subsubsection{Google Cloud Storage}\label{sec:reg-gcs}
Google Cloud also provides native APIs~\cite{gcloudapi} for upload to and download from a storage bucket.
These APIs behave similarly to the \texttt{boto3} that we used for \aws{} and \wasabi{}; we can authenticate once and reuse the credential to transfer all files sequentially for the regression analysis.
\begin{figure}[htb]
\centering
\begin{subfigure}[h]{.49\columnwidth}
\includegraphics[width=\columnwidth,trim=2.5mm 2mm 2mm 2mm,clip]{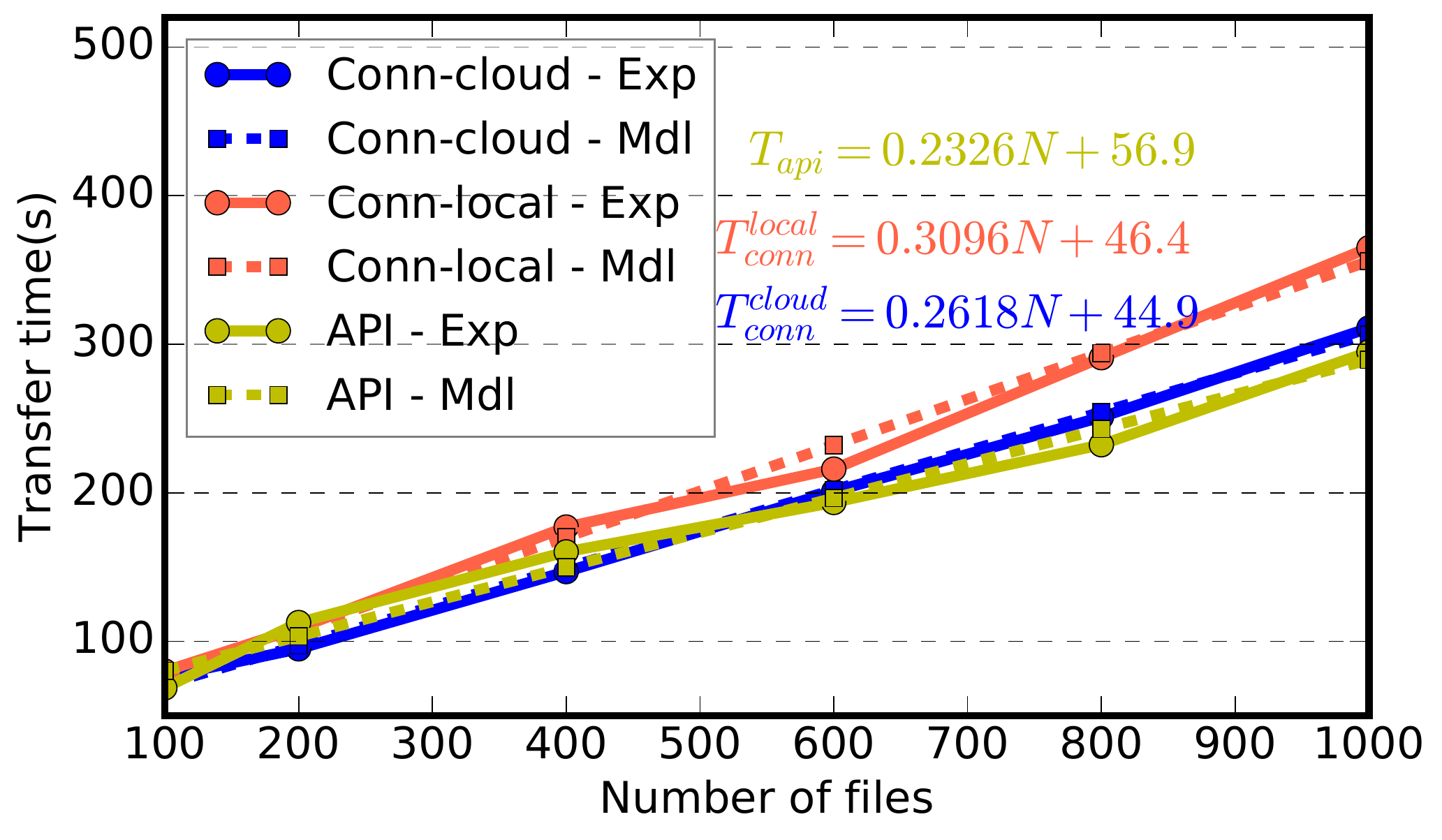}
\caption{Upload}
\label{fig2:gcs-reg-ana-up}
\end{subfigure}
\begin{subfigure}[h]{.49\columnwidth}
\includegraphics[width=\columnwidth,trim=2.5mm 2mm 2mm 2mm,clip]{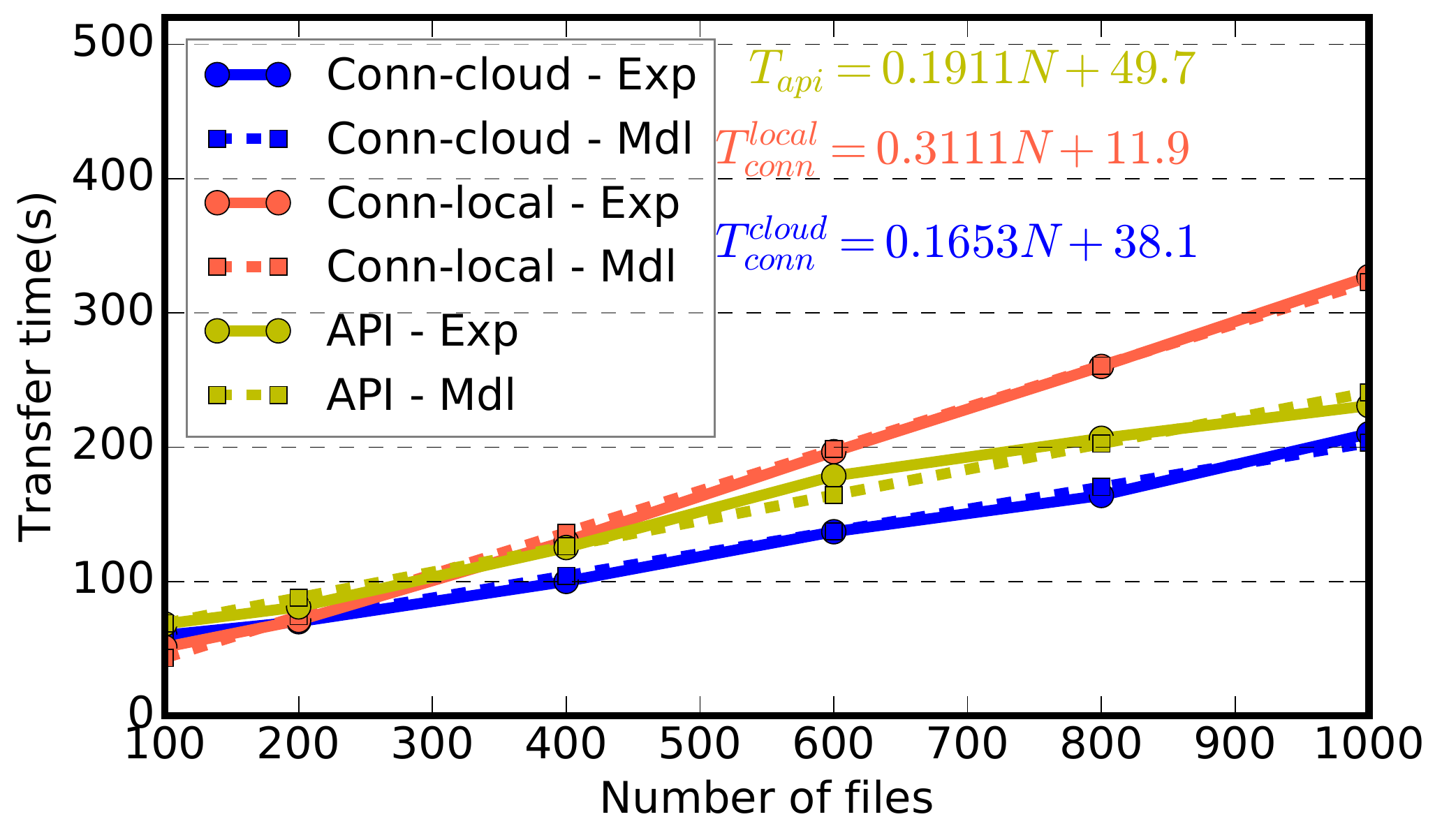}
\caption{Download}
\label{fig2:gcs-reg-ana-down}
\end{subfigure}
\caption{Transfer time vs.\ number of files for 5~GB between the local file system and \gcs{}. \texttt{Conn-cloud}: \gdsi{}
is deployed in the Google Cloud near the bucket (\autoref{fig:dsi-incloud}); \texttt{Conn-local} is deployed in  a science institution (\autoref{fig:dsi-gateway-all}).}
\label{fig:gcs-reg-ana}
\end{figure}
\autoref{fig:gcs-reg-ana} compares experimental results and the fitted performance model.

When the \dsigcs{} is deployed locally, per-file overhead is much higher than that of the native API in both data movement directions. 
As modeled in \autoref{eq:trs-mdl}, however, the \gdsi{} achieves much higher efficiency than the native API does (bias of the linear model is smaller).
In other words, the \gdsi{} will perform better than the API when transferring a few big files but worse when transferring many small files.

In contrast, when the \gdsi{} 
is deployed near the storage bucket on a Google Cloud VM, then GridFTP (optimized for WAN data movement) is used for WAN transfer and the API only within the cloud, namely, to move data between the Cloud VM and storage bucket within the same data center.
As shown in \autoref{fig:gcs-reg-ana}, the per-file overhead of \gdsi{} is slightly worse than that of the API for upload but better than the API for download, thanks to GridFTP WAN data movement optimizations.
Again, the \gdsi{}
achieves much higher efficiency than does the native API.
These results reveal that if 
the \gdsi{}
is deployed near the storage bucket, it will perform better than the native API, if the network bandwidth is not the bottleneck. 

\subsubsection{Google Drive}
Transfers to and from \gdr{} are significantly slower than with the other storage services studied.
Thus, to optimize experiment times and 
 to minimize the influence of external load on our experiments,
we used datasets totaling 1~GB, rather than 5~GB as for other \gdsi{}s, for our regression analysis experiments. 

We see that \dsigdr{} and API perform similarly for uploads from the local file system to \gdr{}.
For downloads, \dsigdr{} introduces a little more per-file overhead than does the native APIs, but its network use efficiency is higher.
Thus, it can achieve similar performance to that of the native API for big files, but it underperforms for smaller files.

\begin{figure}[htb]
\centering
\begin{subfigure}[h]{.49\columnwidth}
\includegraphics[width=\columnwidth,trim=2.5mm 2mm 2mm 2mm,clip]{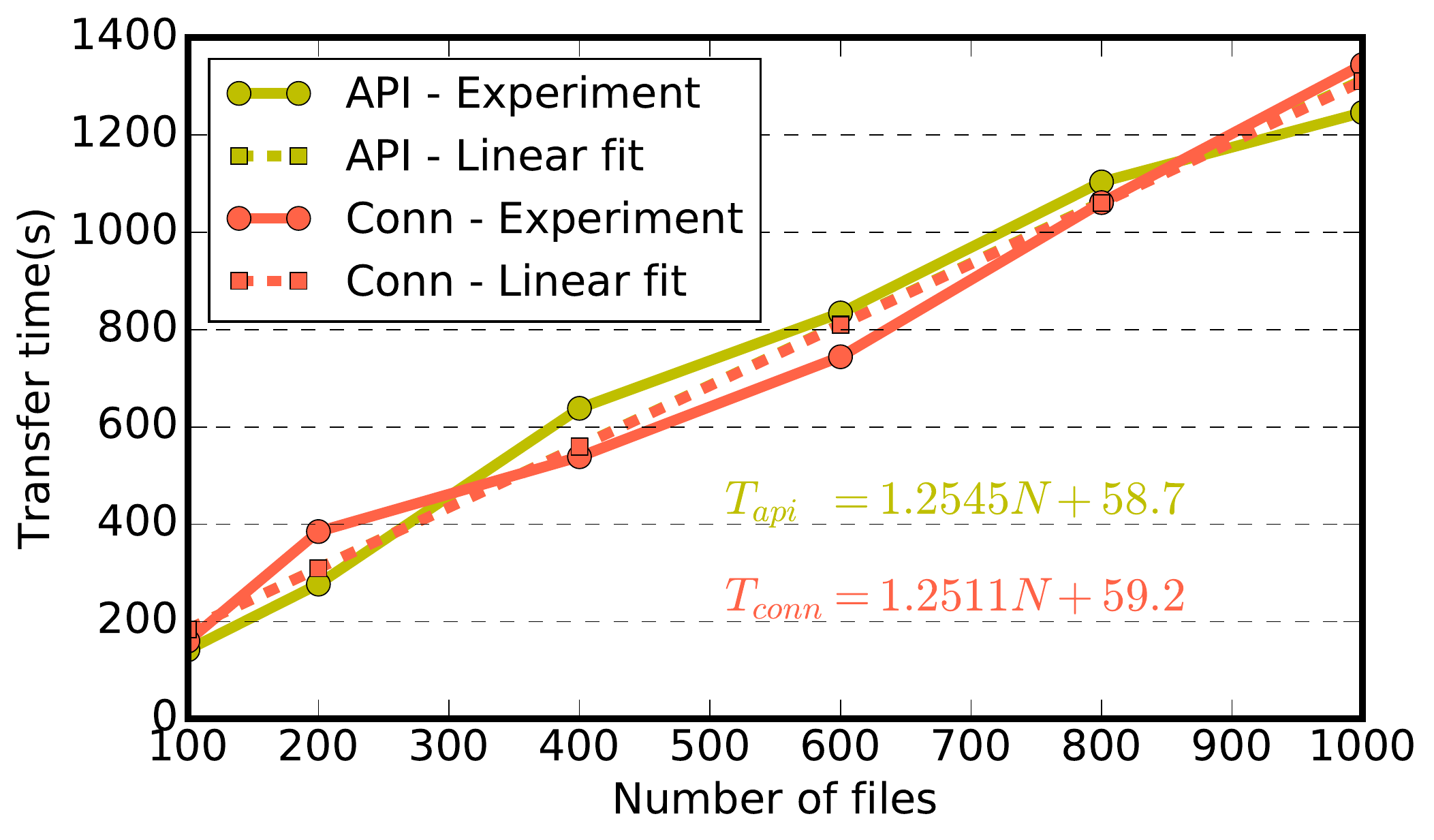}
\caption{Upload to \gdr{}}
\label{fig2:gdr-reg-ana-up}
\end{subfigure}
\begin{subfigure}[h]{.49\columnwidth}
\includegraphics[width=\columnwidth,trim=2.5mm 2mm 2mm 2mm,clip]{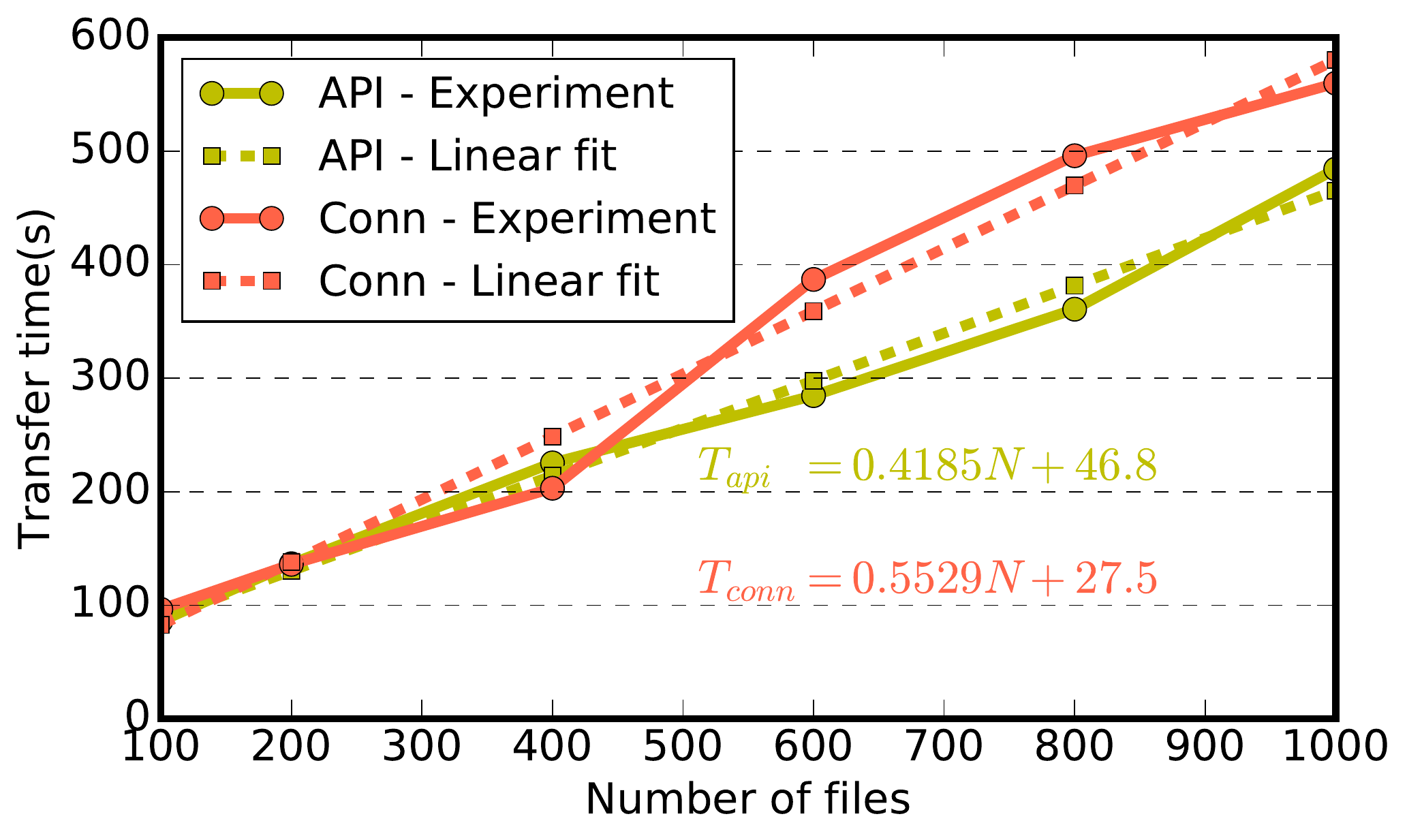}
\caption{Download}
\label{fig2:gdr-reg-ana-down}
\end{subfigure}
\caption{Transfer time vs.\ number of files for 1~GB between local filesystem and \gdr{}}
\label{fig:gdr-reg-ana}
\end{figure}

\subsubsection{Ceph}
Similar to our evaluation for \dsiaws{} and \dsigcs{}, we  consider two deployed scenarios for \dsiceph{}: (1) close to the Ceph storage system (referred to as \texttt{cloud}) and (2) locally in the science institution (referred to as \texttt{local}).

\begin{figure}[htb]
\centering
\begin{subfigure}[h]{.48\columnwidth}
\includegraphics[width=\columnwidth,trim=2.5mm 2mm 2mm 2mm,clip]{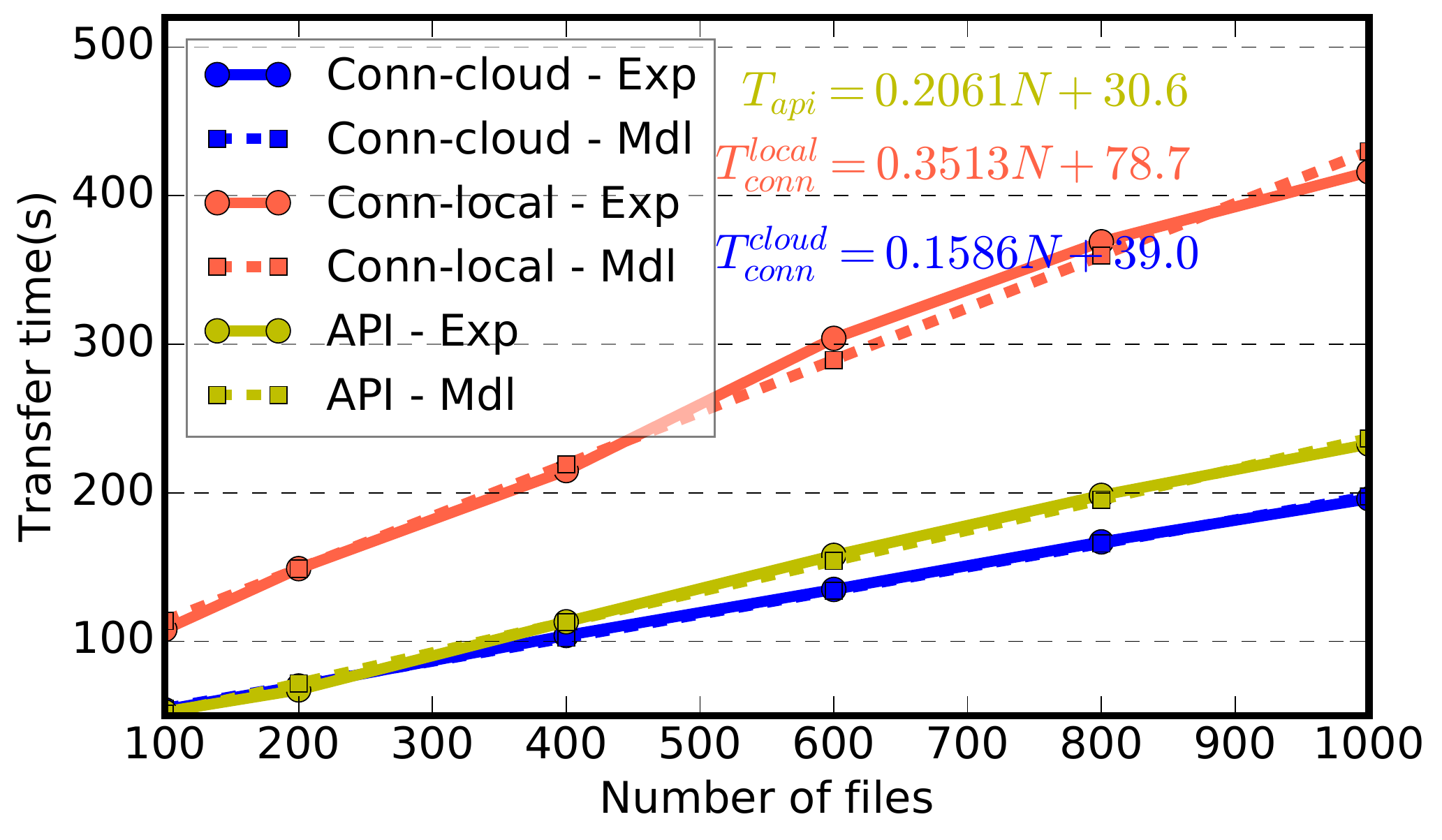}
\caption{Upload to \ceph{}}
\label{fig2:ceph-reg-ana-up}
\end{subfigure}
\begin{subfigure}[h]{.48\columnwidth}
\includegraphics[width=\columnwidth,trim=2.5mm 2mm 2mm 2mm,clip]{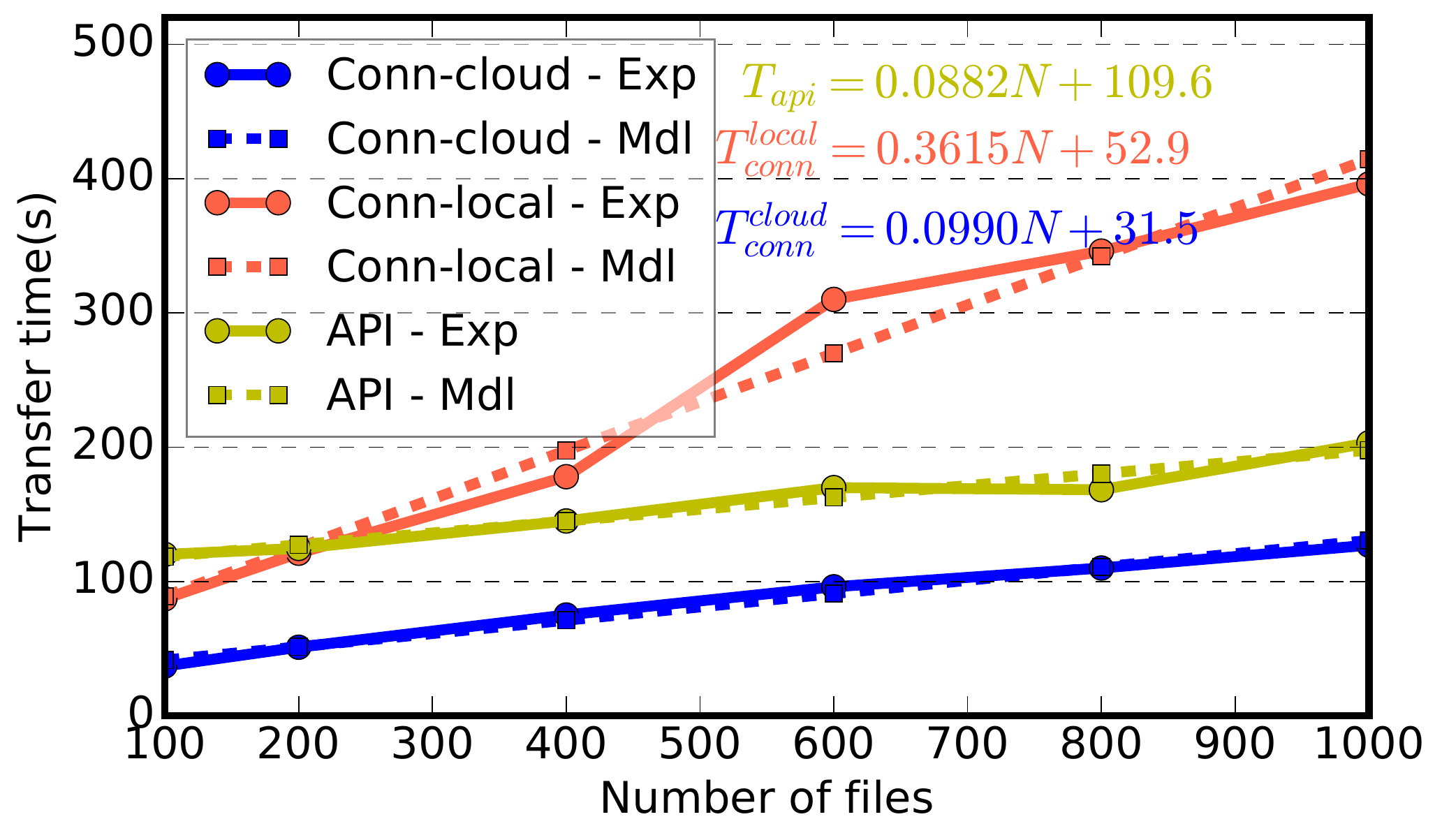}
\caption{Download from \ceph{}}
\label{fig2:ceph-reg-ana-down}
\end{subfigure}
\caption{Transfer time vs.\ number of files for 5~GB between local file system and \ceph{} cloud storage.}
\label{fig:ceph-reg-ana}
\end{figure}

\autoref{fig:ceph-reg-ana} compares the performance model and actual experiment measurement.
We see that, as in the case of \dsiaws{} and \dsigcs{}, the \gdsi{} incurs much lower per-file overheads when deployed near the storage system.
That is mostly because GridFTP allows moving data out-of-order that leads to better efficiency, and GridFTP plays the role to move data over wide-area network when \gdsi{} is deployed near the cloud storage. 

\subsubsection{Box.com}
\begin{figure}[htb]
\centering
\begin{subfigure}[h]{.48\columnwidth}
\includegraphics[width=\columnwidth,trim=2.5mm 2mm 2mm 2mm,clip]{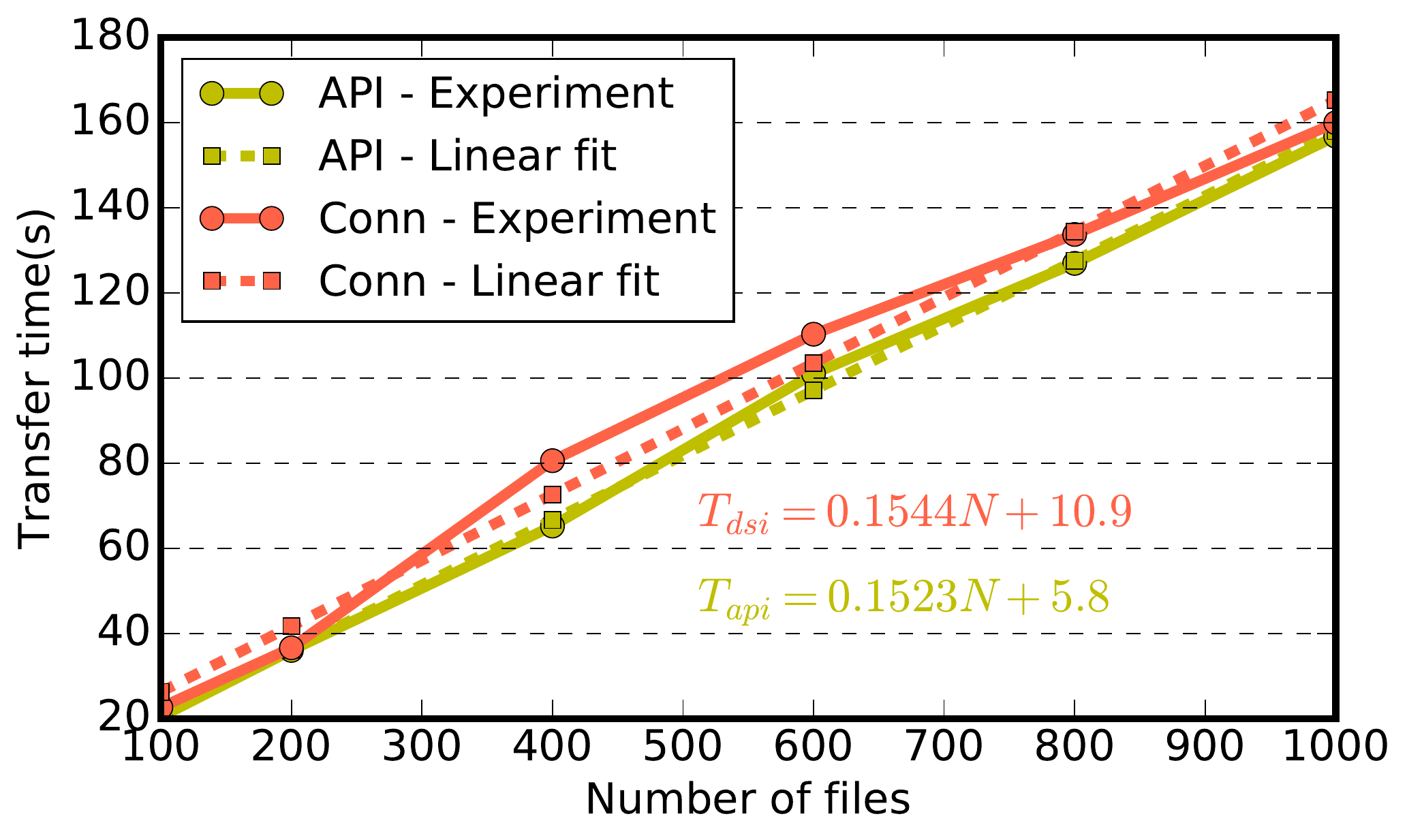}
\caption{Upload to \boxcom{}}
\label{fig2:box-reg-ana-up}
\end{subfigure}
\begin{subfigure}[h]{.48\columnwidth}
\includegraphics[width=\columnwidth,trim=2.5mm 2mm 2mm 2mm,clip]{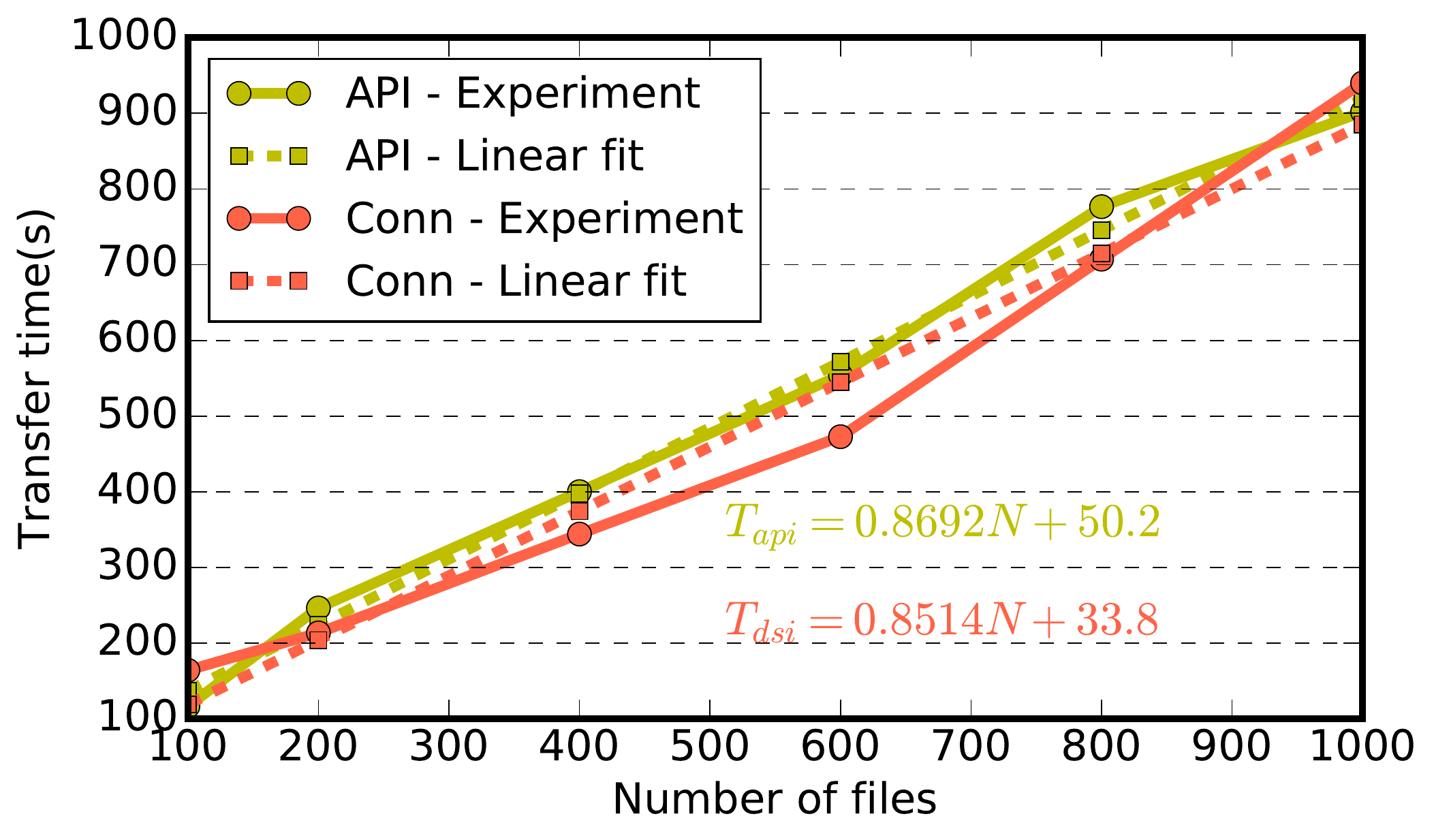}
\caption{Download from \boxcom{}}
\label{fig2:box-reg-ana-down}
\end{subfigure}
\caption{Transfer time vs.\ number of files for 1~GB between the local file system and \boxcom{}}
\label{fig:box-reg-ana}
\end{figure}

As we did for
evaluating other \gdsi{}s, here again we used native APIs (here, those provided by the \textit{Box SDK}~\cite{boxsdk}) to move data between \boxcom{} and local storage in order to compare Box and the \gdsi{}. 
From the experimental measurements in \autoref{fig:box-reg-ana}, we observe that \dsiboxcom{} and the native API have similar per-file overheads. 

\subsection{Transfer Startup Cost}\label{sec:startup-cost}
\begin{figure}[htb]
\centering
\includegraphics[width=.8\linewidth]{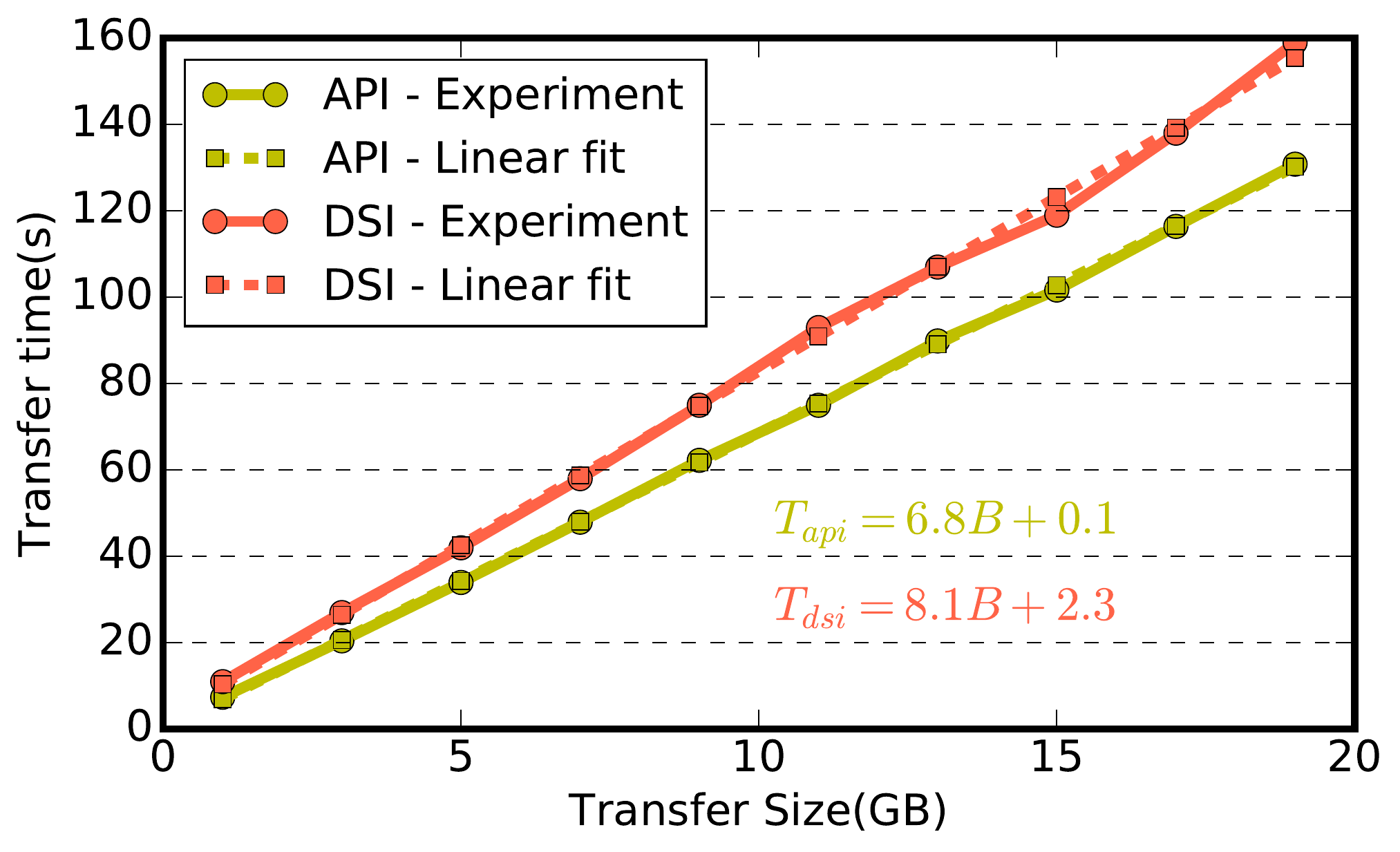}
\caption{Transfer time vs.\ single-file dataset size for upload to \wasabi{}: Globus \gdsi{} third-party and API two-party.}
\label{fig:startup-mdl}
\end{figure}

\autoref{eq:trs-mdl} includes, for each transfer, a startup cost of $S_0$.
This cost varies according to the transfer method used.
If a user logs in to a cloud service and initiates a two-party transfer
directly, the cost may be relatively low.
In the case of a cloud-hosted third-party transfer service such as Globus, it will be higher. 
To measure this cost in different contexts, we 
designed an experiment that transfers a single file of different sizes. 
Thus, the performance model is
\begin{equation}
T = B * t_u + S_0,	
\label{eq:startup-mdl}
\end{equation}
where $B$ is the size of the single file in GB and $t_u$ is the time to transfer 1 GB.
To resolve $S_0$, we transfer a single file with $B\in$\{1, 3, \dots, 17, 19\}~GB from a local file system to a cloud store (in this case, \wasabi{}), and fit the resulting runtimes to \autoref{eq:startup-mdl}.
\autoref{fig:startup-mdl} shows the relation between $B$ and \emph{T}. 
We see a strong linear relationship between $B$ and \emph{T} and a transfer startup cost of 2.3 seconds, which is negligible in most cases except where one transfers a particularly small amount of data in a particularly high-throughput environment.

\section{Throughput Analysis}\label{sec:throughput}
Based on the investigation of per-file overhead, we see that datasets with big files are more friendly to transfer tools~\cite{ccgrid-19}. 
Here, we used the most friendly datasets to benchmark the best transfer performance using different concurrency levels. 
Specifically, in order to use a concurrency of $cc$ with a \gdsi{}, we initiated a transfer with $cc$ files, each of size 1~GB. 
When a native API was used, we initiated $cc$ threads to transfer $cc$ files concurrently.
In practice, aggregated throughput first increases quickly with concurrency and eventually drops slowly, because of local contention. 
As noted in previous studies~\cite{arslan2018big,yildirim2015application,smart-dtn}, there is no one-size-fits-all setting for concurrency. 
Thus, for all experiments in this section, we increased concurrency until we see negative benefit.

\subsection{Wasabi}

\begin{figure}[htb]
\centering
\begin{subfigure}[h]{.49\columnwidth}
\includegraphics[width=\columnwidth,trim=2.5mm 2mm 2mm 2mm,clip]{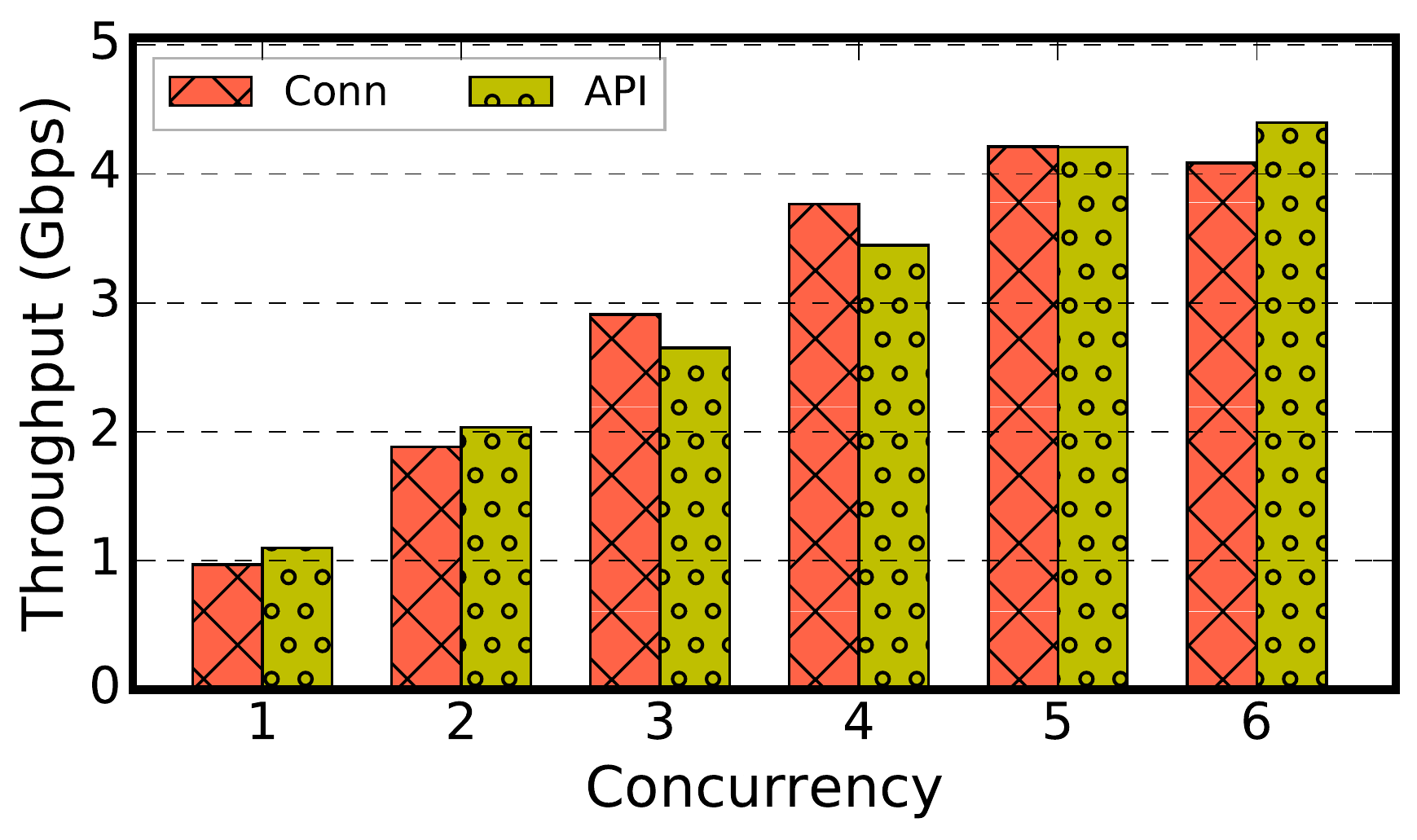}
\caption{Upload to \wasabi{}}
\label{fig2:wasabi-cc2rate-up}
\end{subfigure}
\begin{subfigure}[h]{.49\columnwidth}
\includegraphics[width=\columnwidth,trim=2.5mm 2mm 2mm 2mm,clip]{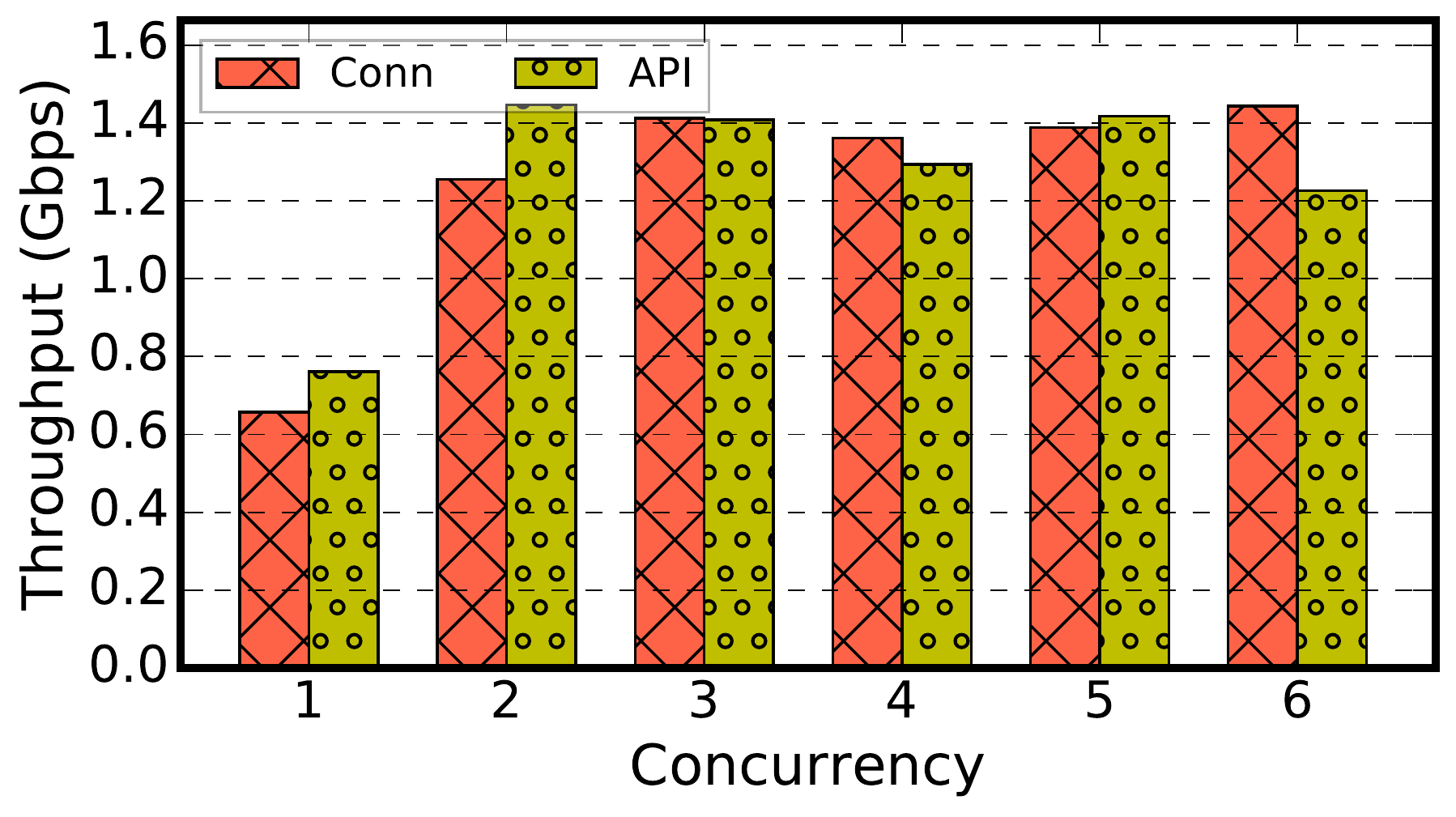}
\caption{Download from \wasabi{}}
\label{fig2:wasabi-cc2rate}
\end{subfigure}
\caption{Transfer performance as a function of concurrency: Globus \gdsi{} third-party and \wasabi{} two-party API}
\label{fig:wasabi-cc2rate}
\end{figure}

\autoref{fig:wasabi-cc2rate} compares the S3 \gdsi{} and Wasabi API.
We see that transferring multiple files concurrently does help to some extent by overlapping the per-file overhead. 
As for throughput, as evaluated in the preceding section, \dsiwasabi{} achieves  performance similar to that of the native API does.

\subsection{AWS S3}
\begin{figure}[htb]
\centering
\begin{subfigure}[h]{.49\columnwidth}
\includegraphics[width=\columnwidth,trim=2.5mm 2mm 2mm 2mm,clip]{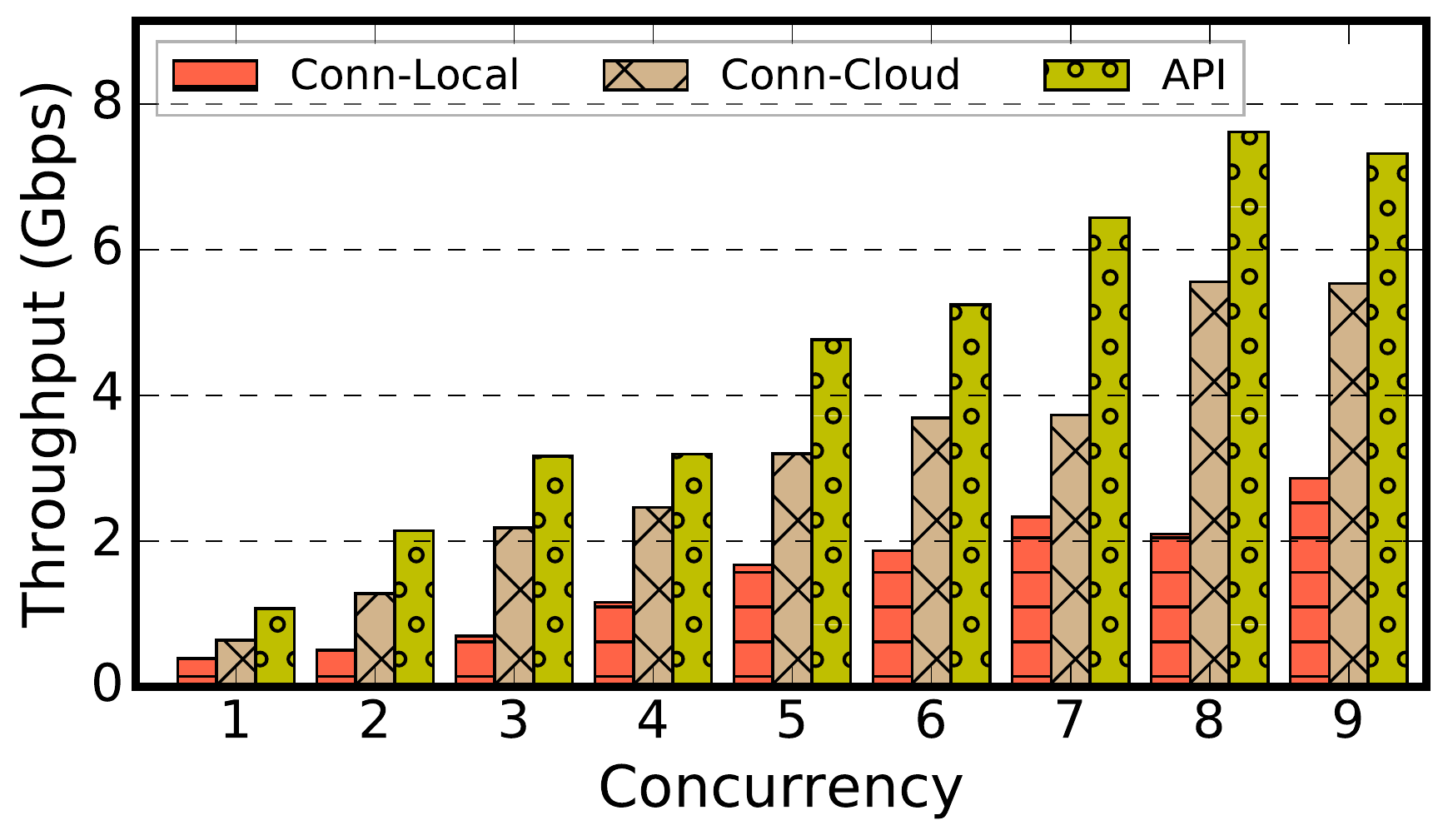}
\caption{Upload to \aws{}}
\label{fig2:aws-cc2rate-up}
\end{subfigure}
\begin{subfigure}[h]{.49\columnwidth}
\includegraphics[width=\columnwidth,trim=2.5mm 2mm 2mm 2mm,clip]{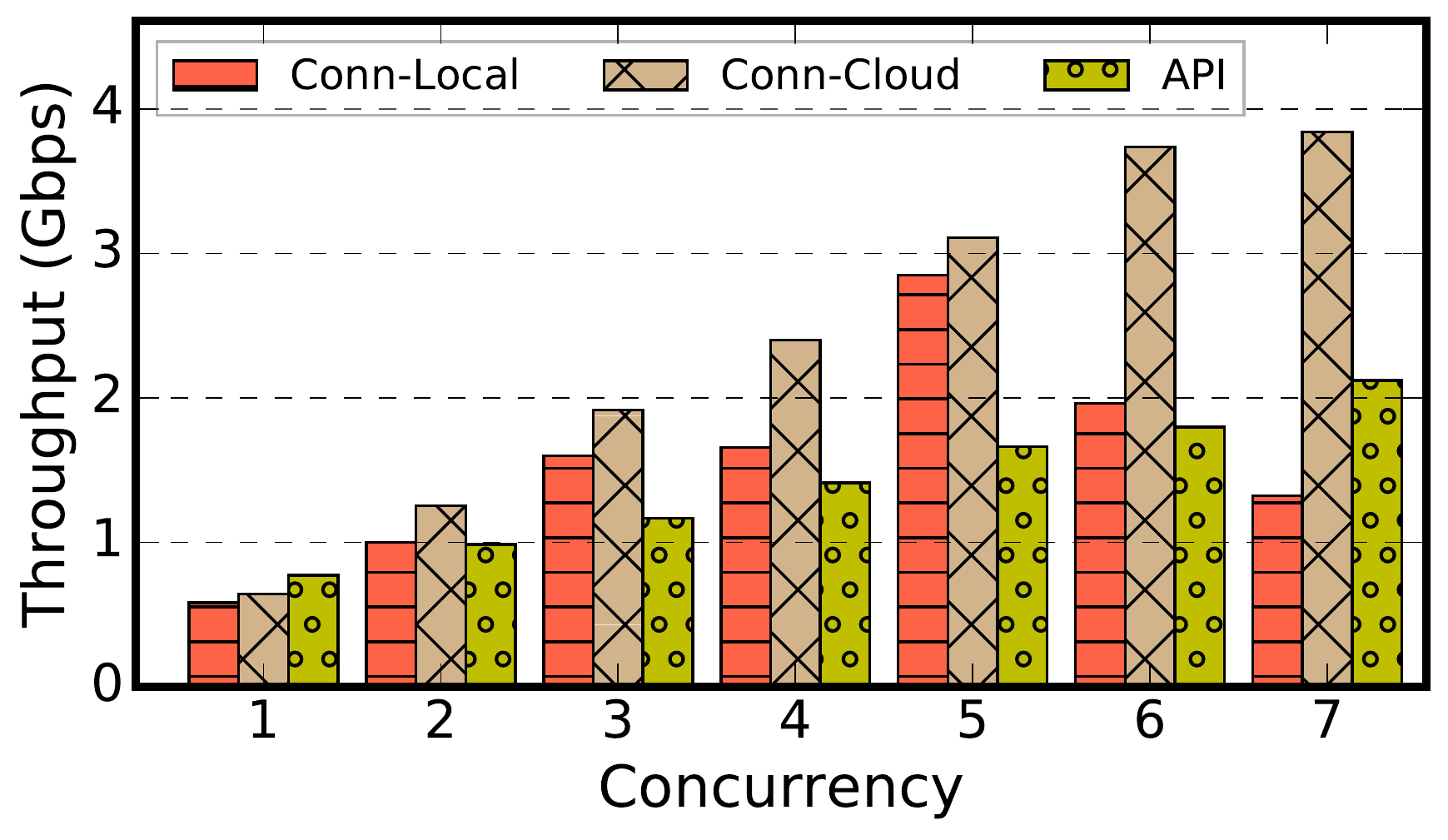}
\caption{Download from \aws{}}
\label{fig2:aws-cc2rate-down}
\end{subfigure}
\caption{Transfer performance as a function of concurrency}
\label{fig:aws-cc2rate}
\end{figure}

\autoref{fig:aws-cc2rate} shows transfer performance between local DTN and AWS S3 as a function of concurrency used. 
We see that uploads to AWS S3 are consistently faster via the AWS API than via \dsiaws{}, while for downloads the reverse is true.
Furthermore, the \gdsi{} performance is consistently better when on AWS rather than local.
We attribute the superior download performance of \dsiaws{} to its use of the wide-area-network-optimized GridFTP, which for example allows out-of-order transmissions. 
Thus, \dsiaws{} can extract data from S3 as fast as S3 will allow via local area network (within AWS region) and transmit them in parallel (out-of-order if needed) over the wide area network using GridFTP. 

For downloads, the limitation seems to be the network performance of the AWS EC2 instance on which the \dsiaws{} is located.
The \texttt{m5.8xlarge} instance (32 vCPU, 128~GB RAM) that we used to host \dsiaws{} on AWS is supposed to deliver 10~Gbps external network performance. However, an \texttt{iperf} test with 16 parallel TCP streams from the AWS instance to our local DTN showed only 4.7~Gbps (i.e., downloads in \autoref{fig2:aws-cc2rate-down}). 

\subsection{Google Cloud Storage}
\begin{figure}[htb]
\centering
\begin{subfigure}[h]{.49\columnwidth}
\includegraphics[width=\columnwidth,trim=2.5mm 2mm 2mm 2mm,clip]{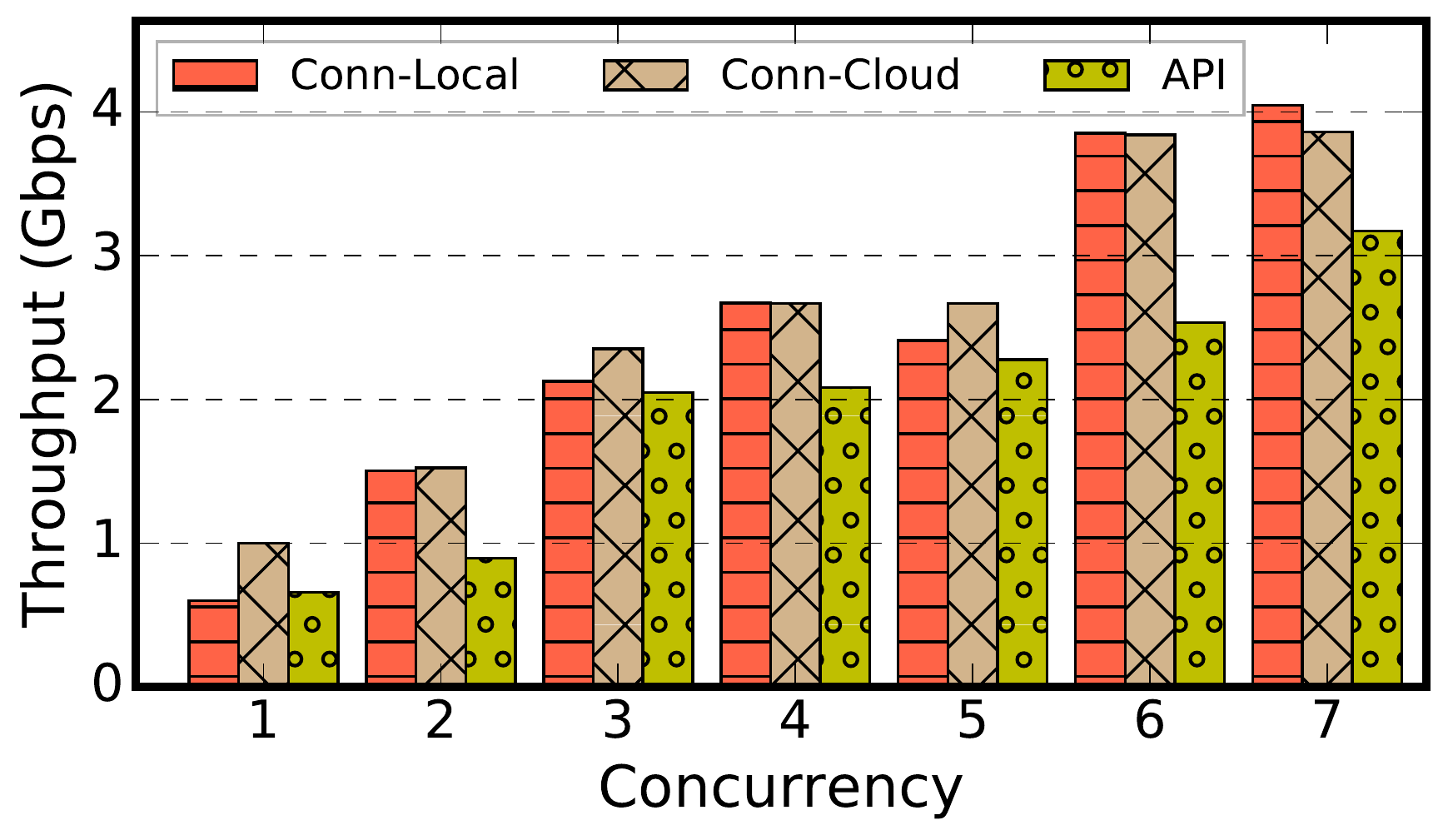}
\caption{Upload}
\label{fig2:gcs-cc2rate-up}
\end{subfigure}
\begin{subfigure}[h]{.49\columnwidth}
\includegraphics[width=\columnwidth,trim=2.5mm 2mm 2mm 2mm,clip]{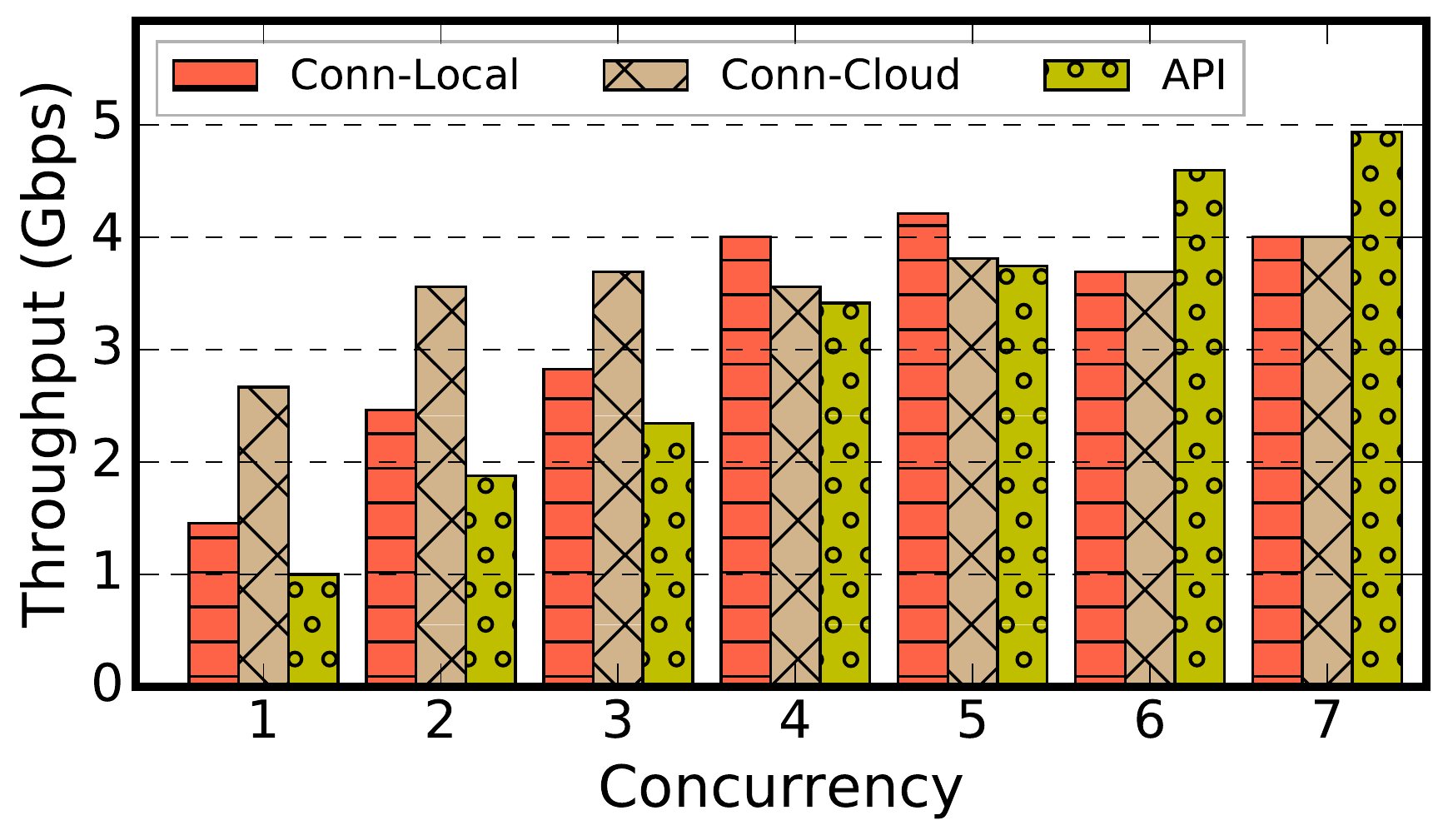}
\caption{Download}
\label{fig2:gcs-cc2rate-down}
\end{subfigure}
\caption{Transfer (upload to and download from \gcs{}) performance as a function of concurrency}
\label{fig:gcs-cc2rate}
\end{figure}

\autoref{fig:gcs-cc2rate} shows transfer performance as a function of concurrency used. In the Conn-cloud case, the \gdsi{} runs on a Google Cloud virtual machine instance with 32 vCPU and 128GB RAM that is close to the \gcs{} bucket. 
We used \texttt{iperf} with 16 parallel TCP streams to measure network bandwidth between our local DTN and the VM instance on Google Cloud; we achieved 4~Gbps from Google Cloud to local DTN (i.e., download) and 7.3~Gbps from local DTN to the Google Cloud instance (i.e., upload). 
Since the data will not go through this VM instance when using the native API, native API transfers are not limited by the above mentioned peak \texttt{iperf} throughput values (which are likely limited by the VM's network). Thus, it is not fair to compare the throughput achieved by the API with that of the throughput achieved by \gdsi{} when the throughput achieved by the API is above the peak \texttt{iperf} measurements.
We see in \autoref{fig2:gcs-cc2rate-up} that the \gdsi{} upload performance is consistently better than that of the native API.

In the download case, since the VM's network egress bandwidth is only 4~Gbps, the comparison after achieving 4~Gbps (there are protocol overheads in practice) does not make sense. 
However, in those experiment that are not limited by network bandwidth (i.e., when concurrency is less than 5), the \gdsi{} clearly performs better than API, with the cloud-placed \gdsi{} (Conn-cloud) performing better than the locally placed connector (Conn-local).
These results are in line with our regression analysis in \S\ref{sec:reg-gcs}.

\subsection{Ceph}
\begin{figure}[htb]
\centering
\begin{subfigure}[h]{.49\columnwidth}
\includegraphics[width=\columnwidth,trim=2.5mm 2mm 2mm 2mm,clip]{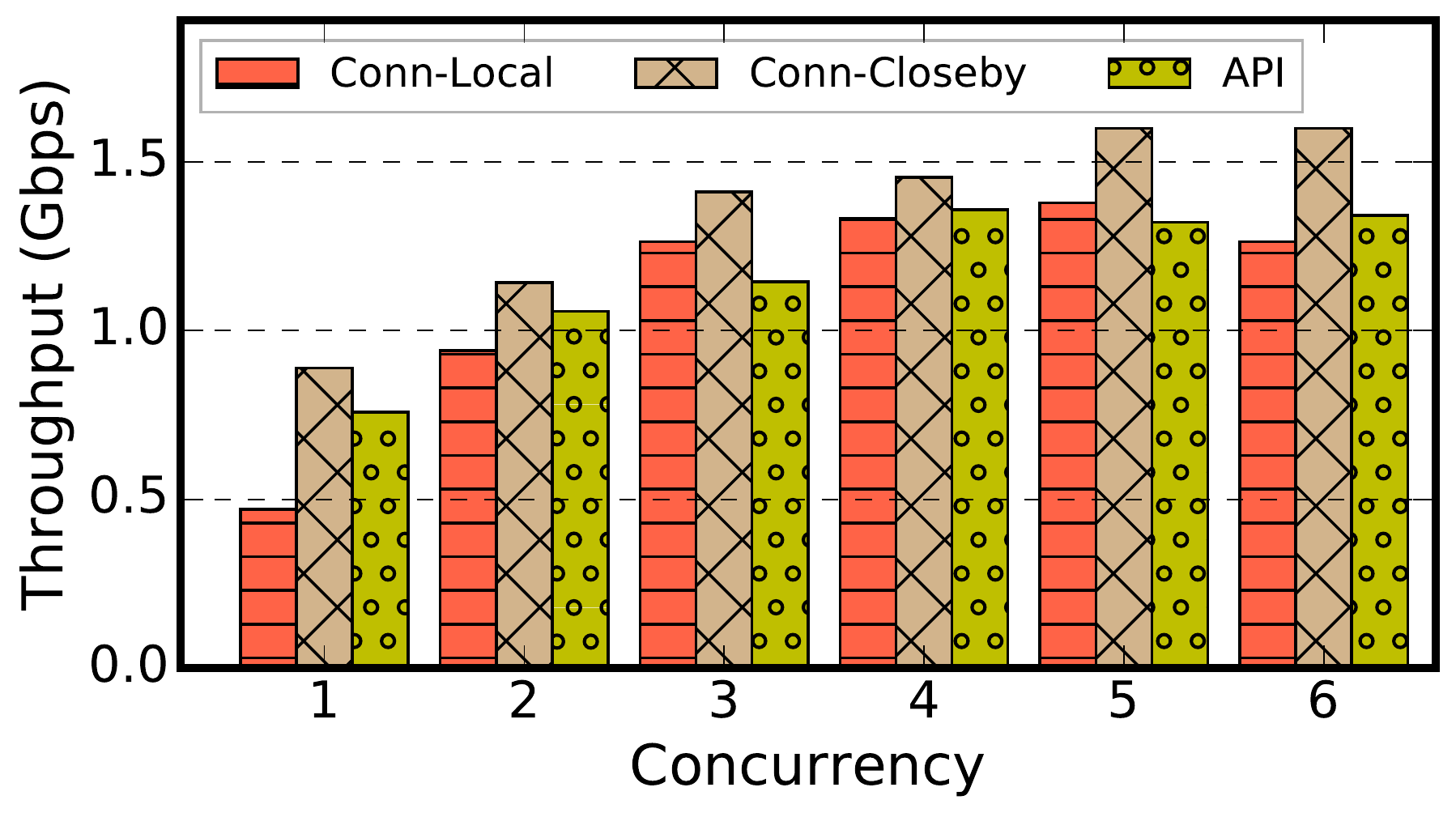}
\caption{Upload}
\label{fig2:ceph-cc2rate-up}
\end{subfigure}
\begin{subfigure}[h]{.49\columnwidth}
\includegraphics[width=\columnwidth,trim=2.5mm 2mm 2mm 2mm,clip]{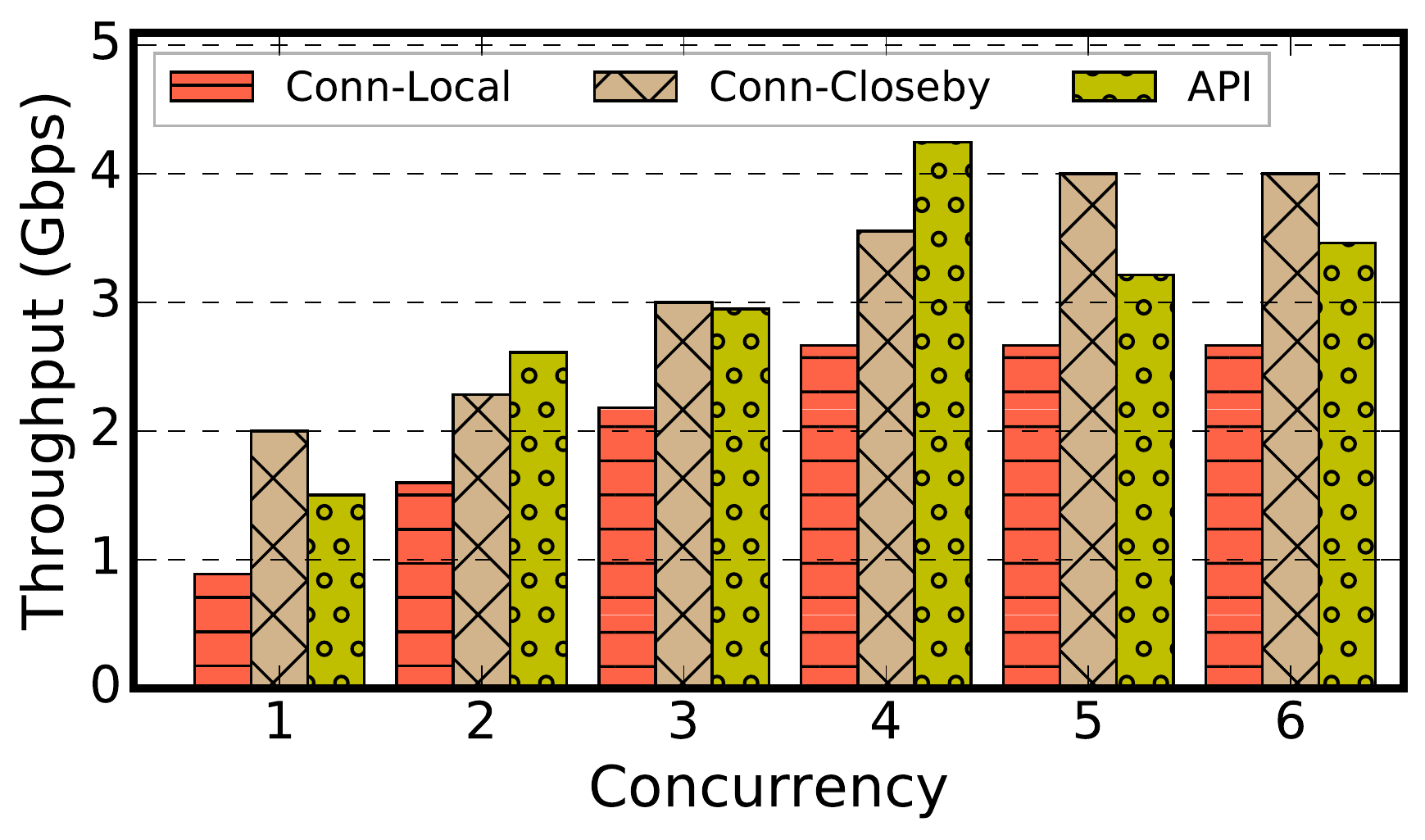}
\caption{Download}
\label{fig2:ceph-cc2rate-down}
\end{subfigure}
\caption{Transfer (upload to and download from \ceph{}) performance as a function of concurrency}
\label{fig:ceph-cc2rate}
\end{figure}

Depending on resource availability and deployment of Ceph storage, similar to \aws{} and \gcs{}, the \dsiceph{} can be deployed in one of two ways: 1) in a local DTN or, 2) near the Ceph storage. 
Here we conduct experiments to benchmark the throughput of \dsiceph{}, and compare with native APIs.
Our Ceph is deployed on a bare metal node at the University of Chicago site of the NSF Chameleon cloud~\cite{chameleon}. 
We deployed \dsiceph{} in two locations: adjacent to the Ceph storage in Chicago and in Texas, at the TACC site of the NSF Chameleon cloud.
Since the data channel of \dsiceph{} uses the S3 protocol, we compared it against using \texttt{boto3} to access Ceph. 
\autoref{fig:ceph-cc2rate} compares the performance of the two \dsiceph{} deployments with that of the native API (i.e., \texttt{boto3}).
\dsiceph{} always get the best performance when deployed near the Ceph system,  
thanks to the optimized data movement over WAN delivered by GridFTP.


\subsection{Inter-cloud Transfers}
The ability to transfer files directly from one cloud store to another, instead of downloading and re-uploading files to and from an intermediate point, such as a user's workstation, can be a major boost to researcher productivity.  
In addition to increasing performance, the fire-and-forget nature of third-party transfer increases reliability and eliminates the need to maintain an intermediate node running for the duration of the transfer from one cloud to another.

\subsubsection{Connector cross-cloud performance}\label{sec2:cross-cloud}
\begin{figure}[htb]
\centering
\begin{subfigure}[h]{\columnwidth}
\includegraphics[width=\columnwidth]{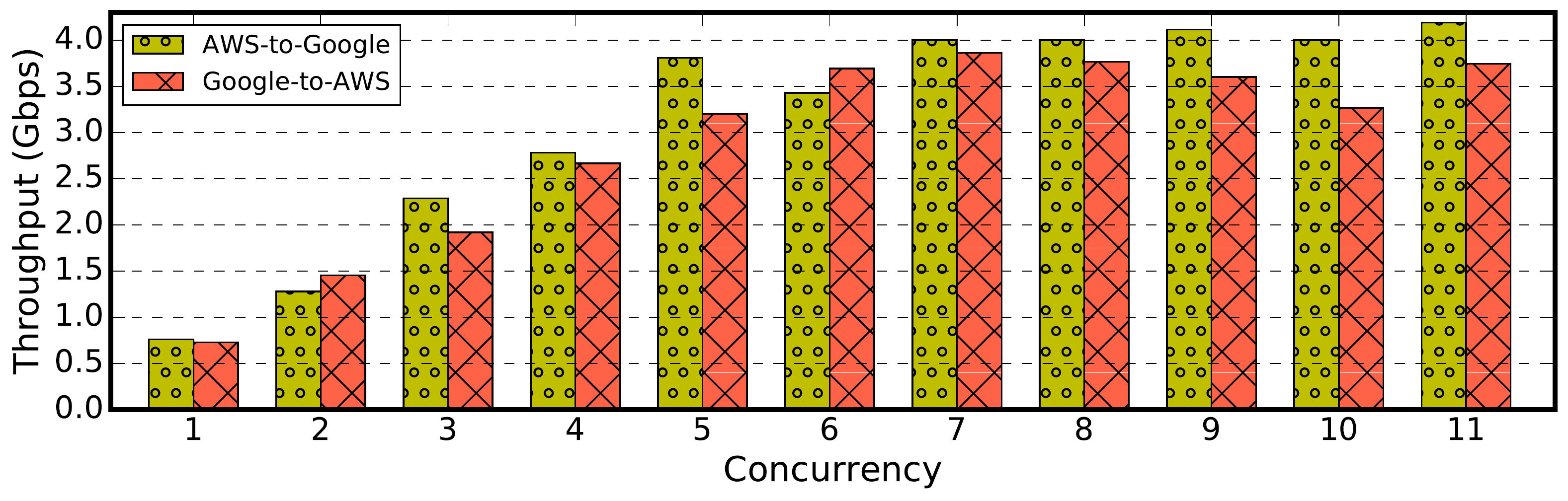}
\vspace{-1ex}
\caption{Via in-cloud DTN}
\vspace{1ex}
\label{fig2:cross-aws-incloud-gcs-cc2rate}
\end{subfigure}
\begin{subfigure}[h]{\columnwidth}
\includegraphics[width=\columnwidth]{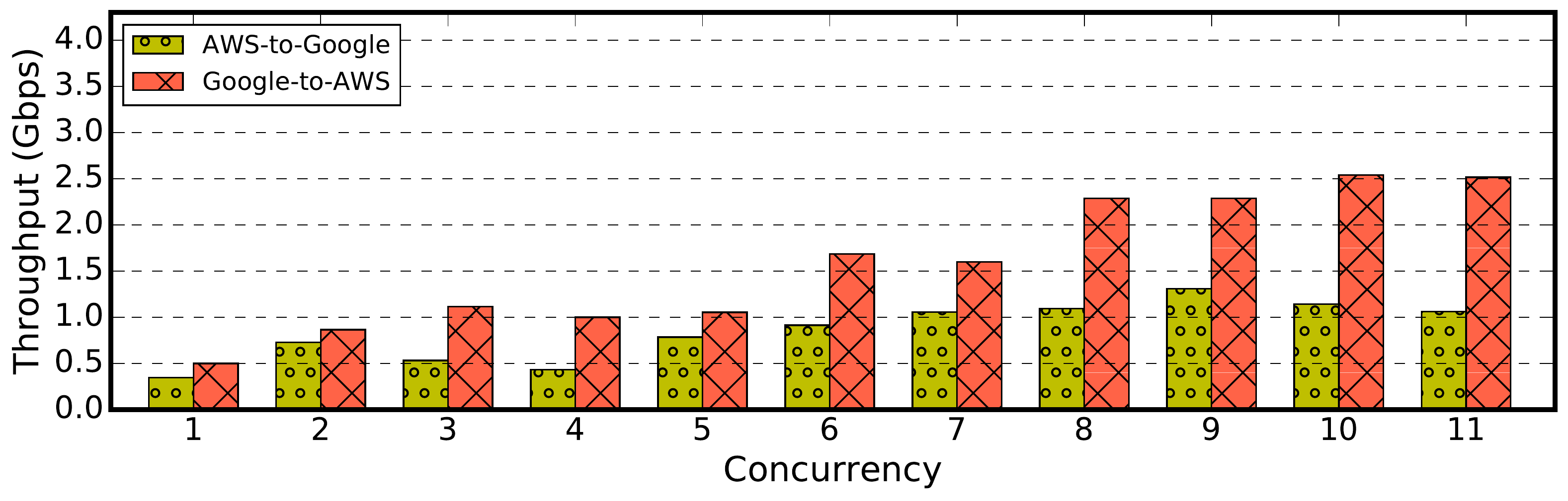}
\vspace{-1ex}
\caption{Via local DTN}
\vspace{1ex}
\label{fig2:cross-aws-cc-gcs-cc2rate}
\end{subfigure}
\caption{Transfer performance between \aws{} and \gcs{} vs.\ concurrency, for local and in-cloud DTN} 
\label{fig:cross-aws-gcs-cc2rate}
\end{figure}

Globus logs show that moving data between \aws{} and \gcs{} is a common use case.
We first evaluated performance for moving data between cloud providers.
\autoref{fig:cross-aws-gcs-cc2rate} shows performance vs.\ concurrency when moving data between \aws{} and \gcs{} using \gdsi{}. Since there is no straightforward or automated way to do cross-cloud transfers using the native cloud storage APIs, we benchmark the performance of \gdsi{} alone to determine best practice for cross-cloud transfers~(more on best practices in \S\ref{sec:best-practice}).
An iperf3 network speed test with 16 parallel TCP streams between the AWS VM and Google Cloud VM achieved about 4.5~Gbps in each direction.
Thus, as shown in \autoref{fig2:cross-aws-incloud-gcs-cc2rate}, \gdsi{} can reach peak throughput when deployed at the cloud provider. 
If deployed locally, however, they achieve only about half of the performance, a reduction that we attribute to network connectivity among AWS, Google Cloud, and the local DTN. 

\subsubsection{Connector comparison}\label{sec:multcloud}
\begin{figure}[htb]
\centering
\includegraphics[width=\columnwidth]{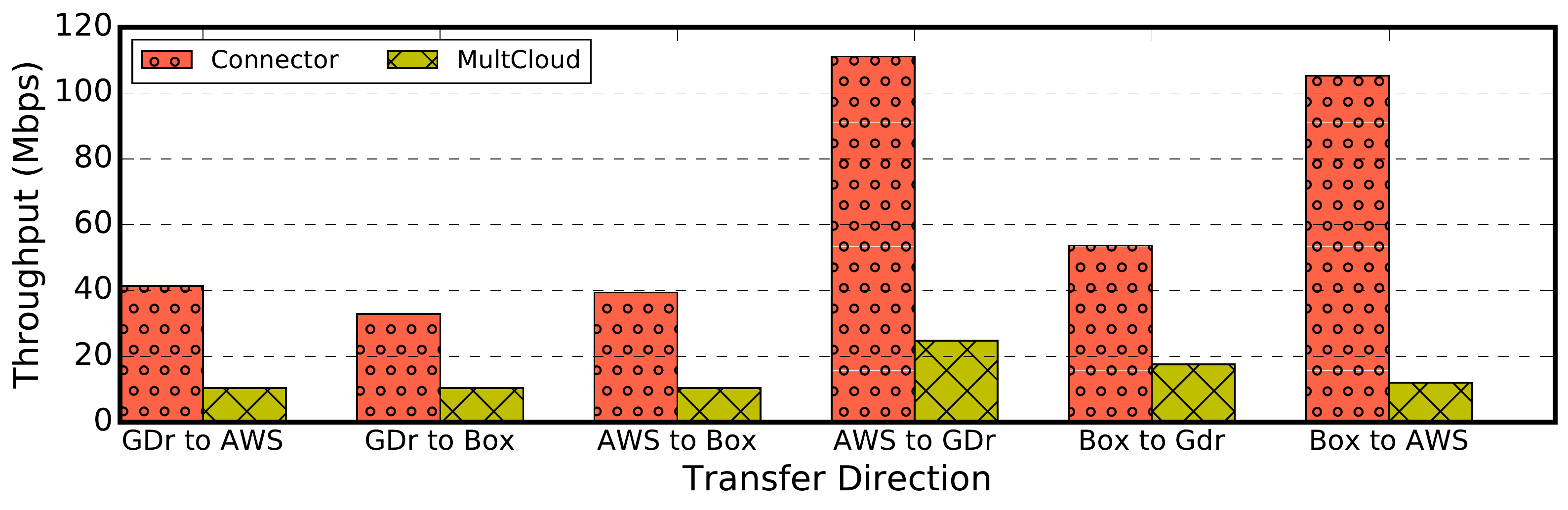}
\caption{Throughput comparison: \texttt{MultCloud} vs.\ Globus}
\label{fig:mc-vs-globus}
\end{figure}

\texttt{MultCloud}~\cite{multcloud}, like Globus, supports data movement across cloud storage services, including
\gdr{}, \boxcom{} and \aws{}. 
In comparing performance, we used our analysis of file size characteristics~\cite{hpdc18-zliu} to select a test dataset of 50 files totaling 1~GB. 
Since the free trial version of \texttt{MultCloud} only supports transferring files one by one, we also set concurrency to one for the Globus \gdsi{} for the experiment.
We used a local DTN to run the corresponding \gdsi{} for the experiment, although a cloud-based DTN gives better performance. 
We see in
\autoref{fig:mc-vs-globus} that
the \gdsi{} outperforms \texttt{MultCloud} in all cases. 

\section{Integrity Checking}\label{sec:integrity}
It is good practice to perform integrity checking on files transferred over wide area networks because factors such as faulty routers and file systems can cause silent data corruption~\cite{stone2000crc,hpdc18-zliu,charyyev2019towards}.
The 16-bit TCP checksum is inadequate to catch network transmission errors, and other errors can occur when accessing storage.
Indeed, a recent study~\cite{hpdc18-zliu} reported at least one checksum failure per 1.26 TB moved from storage to storage over a wide area network.
While this number is likely an over-estimate, 
as it does not distinguish between data corruption and cases in which a file is modified while a transfer is in progress, it emphasizes the importance of integrity checking. 

\begin{figure}[htb]
\centering
\includegraphics[width=\linewidth]{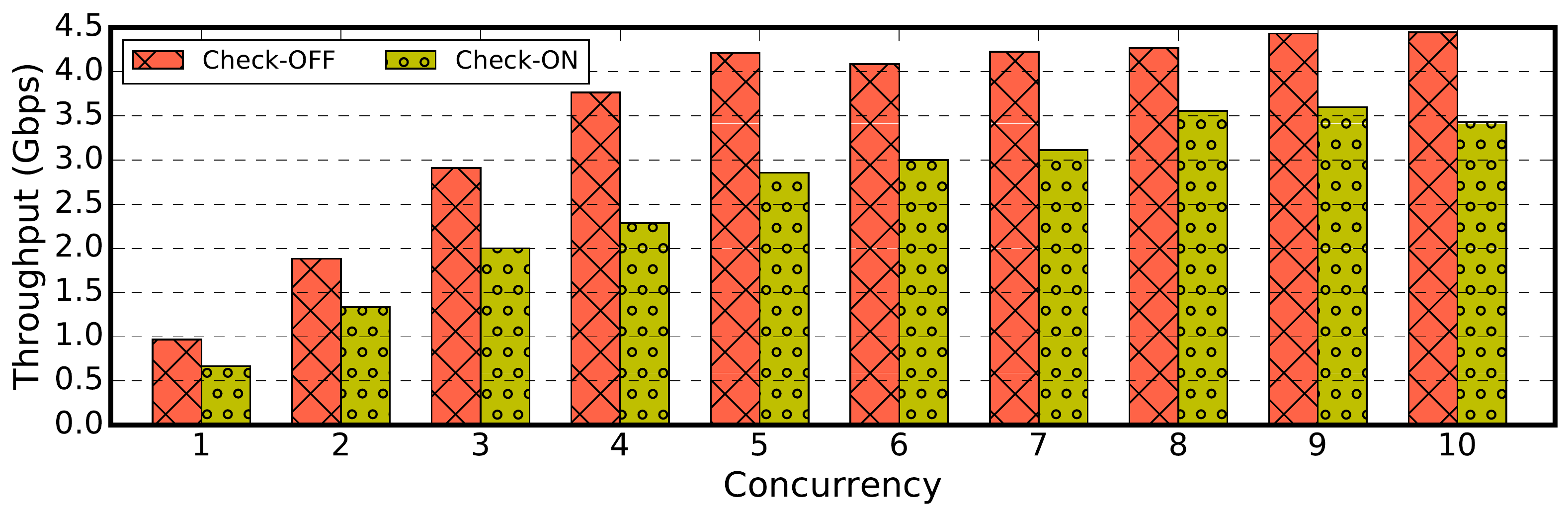}
\caption{Transfer (upload to \wasabi{}) performance, with integrity checking ON versus OFF}
\label{fig:wasabi-cc2rate-up-check}
\end{figure}

\begin{figure}[htb]
\centering
\includegraphics[width=\linewidth]{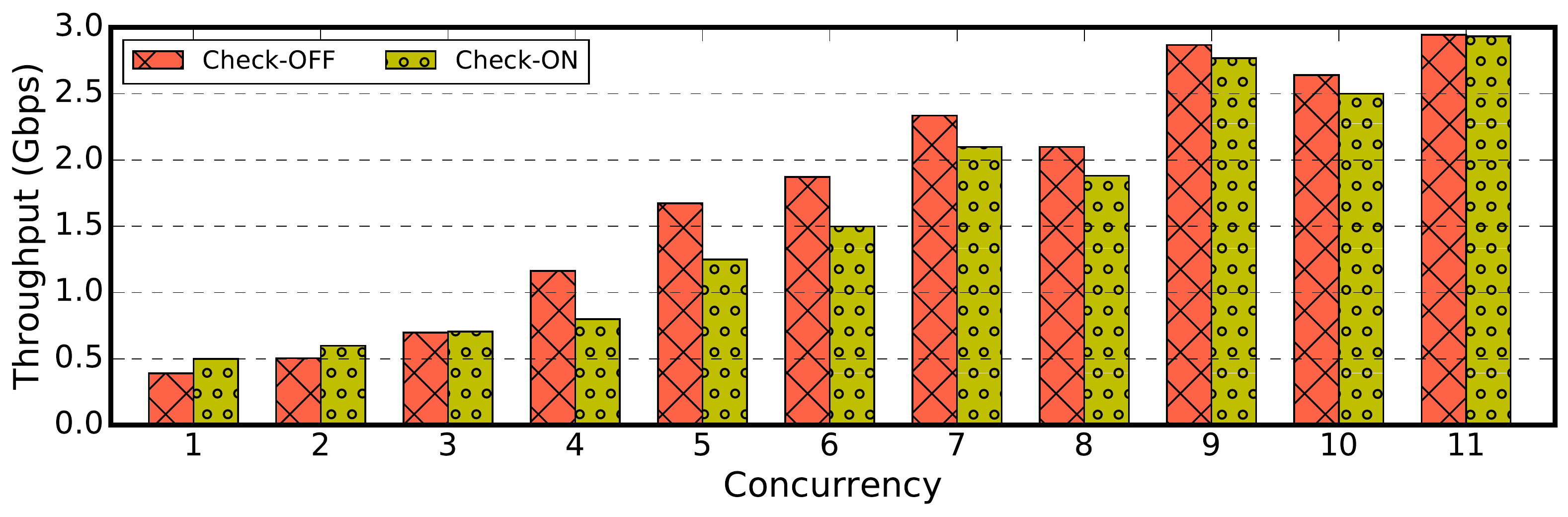}
\caption{Transfer (upload to \aws{}) performance, with integrity checking ON versus OFF}
\label{fig:aws-cc2rate-up-check}
\end{figure}

\begin{figure}[htb]
\centering
\includegraphics[width=\linewidth]{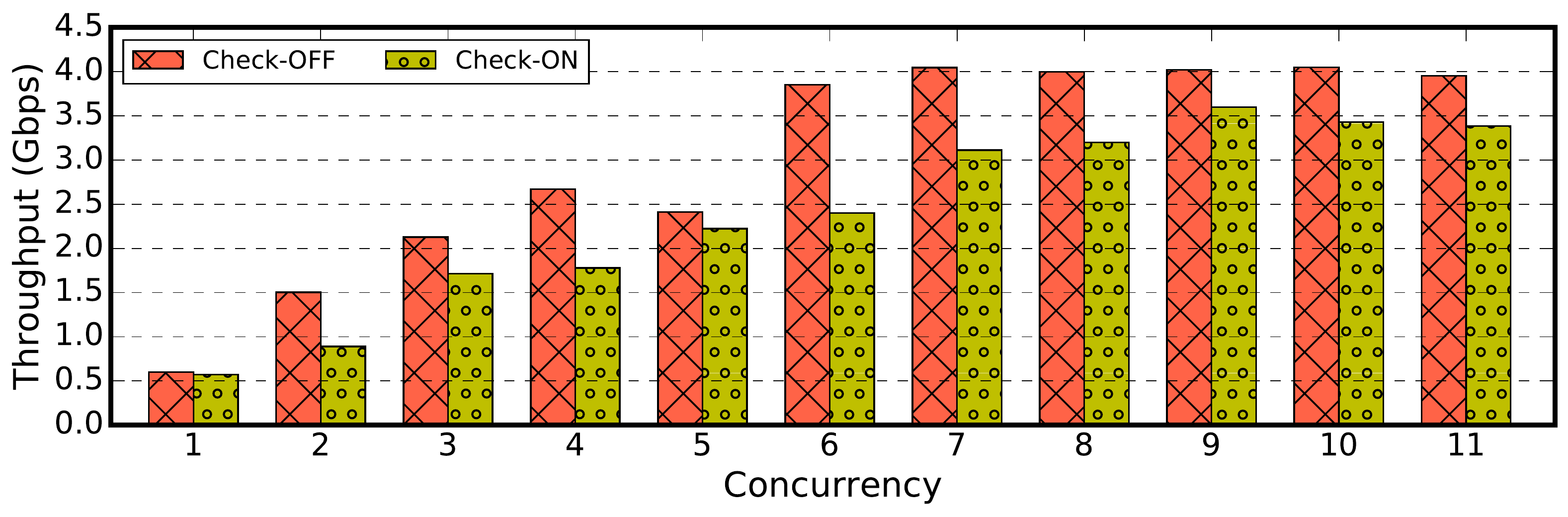}
\caption{Transfer (upload to \gcs{}) performance, with integrity checking ON versus OFF}
\label{fig:gcs-cc2rate-up-check}
\end{figure}

The \gdsi{} abstraction interface supports integrity checking via GridFTP~\cite{globus-gridftp}. 
A client can verify transmission integrity by having a file read and a checksum computed at the source before transmission and then reread and a second checksum computed at the destination. 
This ``strong integrity checking'' approach has the advantage that it can detect not only errors incurred during data transport over the network but also errors incurred while writing data.
However, the additional read operations can impact performance, particularly if a \gdsi{} is located remotely from cloud storage.
Given the wide variety of storage systems, \gdsi{} placement strategies, and transfer workloads, we cannot provide a complete analysis of integrity checking costs. 
However, we present some relevant results for high throughout storage systems (where even a small integrity checking overhead can have a significant influence) in~\autoref{fig:wasabi-cc2rate-up-check}--\autoref{fig:gcs-cc2rate-up-check} for \wasabi{}, \aws{} and \gcs{} respectively.
In each case, the \gdsi{} is located on a computer in our institution(Argonne), and the transfer involves $c$ 300~MB files, where $c$ is the concurrency. 
As one can see, transfer rates are lower when integrity checking is enabled, but not remarkably so, given that the file is being reread over the wide area network after writing.

\section{Best Practice}\label{sec:best-practice}
The GridFTP-based Globus transfer service has been heavily optimized for moving data over wide area networks~\cite{pb-a-day, globus-gridftp}. Here we provide recommendations for \gdsi{} deployment when aiming either to maximize throughput or to minimize costs.

\subsection{Throughput Maximization}
Best practice for throughput maximization when moving data to and from cloud storage is to deploy the corresponding \gdsi{} near the cloud storage service. 
This means, for example, using a Google Cloud computing instance as a DTN to run one or more \dsigcs{} and using AWS EC2 instance(s) to run one or more \dsiaws{}.
Moreover, benchmark experiments in \S\ref{sec2:cross-cloud} also shows that for inter-cloud transfers, this configuration (deploying \gdsi{} near the cloud storage) can achieve a 100\% improvement in throughput compared to the configuration in which \gdsi{} is deployed locally at users' site (or at a location that is not closer to the cloud storage).
The transfer throughput achievable in these two cases depends on the size (in terms of vCPUs and memory) of the allocated instance(s).
We have found that two vCPUs and 4~GB of memory are needed to saturate a 10~Gbps network.
In order to achieve high performance with reduced cost, these cloud-hosted DTN instances can adopt an elastic resource allocation approach, increasing resources allocated to the \gdsi{} when demand is high and reducing it at other times~\cite{elastic-DTI}. 

We note that such cloud-hosted \gdsi{} can be shared by several science institutions that use the same federated authentication mechanism, such as XSEDE~\cite{towns2014xsede}. 

\subsection{Cost Minimization}
An alternative deployment approach is to run \gdsi{} on computers hosted at science institutions. 
This approach does not require any additional hardware, but it means that all accesses to cloud storage involve data transfers with cloud provider protocols.
The results presented earlier in this paper suggest that this approach will lead to little performance loss for datasets with large files but significant performance loss for datasets with many small files.  

Performance-cost calculations may be different when using integrity  checking, as discussed in the next section.
Since \gdsi{} integrity checking involves rereading a file after writing and since cloud storage providers usually charge for network usage when data is moved out of the cloud, it is advantageous when integrity checking is enabled to deploy a cloud storage \gdsi{} in the same cloud as the storage.

\section{Related Work}\label{sec:related-work}
Others have developed implementations of the Globus GridFTP DSI. EUDAT~\cite{van2015d5} implemented a DSI~\cite{b2stage} for the Integrated Rule-Oriented Data System (iRODS)~\cite{rajasekar2010irods} data management software.
S{\'a}nchez et al.~\cite{sanchez2006parallel} proposed a parallel DSI for GridFTP and offered an implementation for the MAPFS~\cite{PEREZ2006620} parallel file system. 
A DSI implementation for OpenStack Object Storage (Swift) is also available~\cite{swift-dsi}.
However, no DSI implementation targets cloud stores, and none provide performance evaluations.

Others have developed uniform interfaces to cloud storage, but by supporting multiple protocols in a client, not a \gdsi{} as proposed here.
We described MultCloud in \S\ref{sec:multcloud}.
Rclone~\cite{rclone} is a command line program that offers a rsync-like tool to synchronize files for cloud storage. It integrates APIs for various cloud stores but does not provide transfer management functionality.
iRODS implements an interface to AWS S3~\cite{wan2009integration}. 

As cloud computing has become the de facto standard for big data processing, Abramson et al.~\cite{Abramson2019} proposed the Metropolitan Data Caching Infrastructure~(MeDiCI) architecture to simplify the movement of data between different clouds and a centralized storage site. 
It is similar as the scenario we evaluated in \S\ref{sec2:cross-cloud} but MeDiCI is cache-based target at on-demand cloud computing.

Liu et al.~\cite{ccgrid-19} used regression analysis to measure per-file overhead indirectly, and concluded that the bottleneck in transferring many small files between HPC facilities is not any single subsystem but rather the per-file overheads introduced by the major components in wide area file transfers. Deelman et al.~\cite{panorama2017} have developed similar models.
The benefits of parallel streams for transfer performance are well known~\cite{hacker2004improving}.
Several researchers have studied the impact of concurrency, parallelism, and other parameters on GridFTP transfer performance~\cite{arslan2018big,yildirim2015application,smart-dtn,arslan2016harp,arslan2018high}, for example, based on historical data~\cite{arslan2018high} or lightweight probing~\cite{arslan2016harp}.

\section{Conclusion}\label{sec:conclusion}
In this paper, we described an architecture, interfaces, and implementation methods for unifying the interface to a wide range of storage systems.  
This architecture enables the plug-and-play integration of storage \gdsi{} for different storage systems that simplifies both the use of different storage systems and the development of new \gdsi{}.
Integration of these \gdsi{} with the Globus data transfer service enables data movement across various storage systems in a ``fire-and-forget" fashion. 
We described \gdsi{} implementations for a range of storage types, from POSIX file systems to HPC parallel file systems and cloud object stores.
We used a performance-model-based analysis to evaluate \gdsi{} implementations and used both experiments and analysis to draw conclusions about implications for the design and implementation.
The proposed performance evaluation method can also be used for inspecting and explaining performance of any other file transfer services.  
We conclude that the \gdsi{} model enables effective use of distributed storage with only modest performance loss relative to native in most cases---and performance improvements in other cases, due to the optimization of data movement over wide area networks delivered by the open source GridFTP.

\section*{Acknowledgments}
This material was based upon work supported by the U.S. Department of Energy,Office of Science, under contract DE-AC02-06CH11357.

\bibliographystyle{ACM-Reference-Format}
\bibliography{globus-dsi}

\section*{Government License}
The submitted manuscript has been created by UChicago Argonne, LLC, Operator of Argonne National Laboratory (“Argonne”). Argonne, a U.S. Department of Energy Office of Science laboratory, is operated under Contract No. DE-AC02-06CH11357. The U.S. Government retains for itself, and others acting on its behalf, a paid-up nonexclusive, irrevocable worldwide license in said article to reproduce, prepare derivative works, distribute copies to the public, and perform publicly and display publicly, by or on behalf of the Government.  The Department of Energy will provide public access to these results of federally sponsored research in accordance with the DOE Public Access Plan. http://energy.gov/downloads/doe-public-access-plan.
\end{document}